\documentclass[11pt]{article}
\usepackage{graphicx,latexsym}

  \def\pd{\partial} \def\pp{\prime} \def\a{\alpha} \def\b{\beta} \def\dl{\delta} \def\s{\sigma}  \def\vphi{\varphi} \def\eps{\epsilon} 
 \def\lam{\lambda} \def\Lam{\Lambda} \def\gm{\gamma} \def\Gm{\Gamma}
 \def\Om{\Omega} \def\nb{\nabla} \def\sq{\sqrt} \def\e{\hbox{\large \it e}}
\def\half{\frac{1}{2}} \def\fr{\frac}  

\def\P{{\rm P}} \def\QG{{\rm QG}} \def\pl{{\rm pl}} \def\M{{\rm M}}

\def\hphi{{\hat \phi}}

\def\bnb{{\bar \nabla}} \def\bg{{\bar g}} \def\bDelta{{\bar \Delta}} \def\bR{{\bar R}}
\def\bE{{\bar E}}    

\def\bT{{\bf T}}  \def\bx{{\bf x}} \def\by{{\bf y}} \def\bk{{\bf k}} 
\def\bp{{\bf p}} \def\bq{{\bf q}}     \def\D{{\bf D}}

\def\lap3{~| \!\!\! \partial^2} \def\dlap3{~| \!\!\! \partial^4}

\def\barH{{\bar H}}  \def\barY{{\bar Y}}  \def\barZ{{\bar Z}}

\begin{document}

\begin{center}
{\large {\bf Study of Nonlinear Evolution of Spacetime Fluctuations in Quantum Gravity Inflation for Deriving Primordial Spectrum}} \\ 
\end{center}

\begin{center}
{\sc Ken-ji Hamada}
\end{center}

\begin{center}
{\it Institute of Particle and Nuclear Studies, KEK, Tsukuba 305-0801, Japan  \\ and Graduate Institute for Advanced Studies, SOKENDAI, Tsukuba 305-0801, Japan}
\end{center}

\begin{abstract}
\noindent
We study the evolution of quantum fluctuations of gravity around an inflationary solution in renormalizable quantum gravity, in which the initial scalar-fluctuation dominance is shown by the background-free nature expressed by a special conformal invariance. Inflation ignites at the Planck scale and continues until spacetime phase transition occurs at a dynamical scale about $10^{17}$GeV. We can show that during inflation, the initially large scale-invariant fluctuations reduce in amplitude to the appropriate magnitude suggested by tiny CMB anisotropies. The goal of this research is to derive the spectra of scalar fluctuations at the phase transition point, that is, the primordial spectra. A system of nonlinear evolution equations for the fluctuations is derived from the quantum gravity effective action. The running coupling constant is then expressed by a time-dependent average following the spirit of the mean field approximation. In this paper, we determine and examine various nonlinear terms, not treated in previous studies such as the exponential factor of the conformal mode. These contributions occur during the early stage of inflation when the amplitude is still large. Moreover, in order to verify their effects concretely, we numerically solve the evolution equation by making a simplification to extract the most contributing parts of the terms in comoving momentum space. The result indicates that they serve to maintain the initial scale invariance over a wide range beyond the comoving Planck scale. This is a challenge toward the derivation of the precise primordial spectra, and we expect  in the future that it will lead to the resolution of the tensions that have arisen in cosmology.
\end{abstract}

\section{Introduction and Summary}
\setcounter{equation}{0}

\noindent
Inflation is an unbelievable idea proposed by Guth \cite{guth}, Sato \cite{sato}, and Starobinsky \cite{starobinsky} to solve the horizon and flatness problems, which claims that there was a period of exponential expansion in the universe before the big bang.\footnote{
See also the historic works of E. B. Gliner, which is memorialized in \cite{yk}.} 
Remarkably, if we genuinely embrace this concept, a significant portion of the universe we observe today originated from a region even smaller than the Planck length.

Going over the Planck-scale wall requires quantization of gravity with not only renormalizability but also background freedom. Defying the common perception in this research area, we have proposed asymptotically background-free quantum gravity \cite{hs, hamada02, hamada14re, hm16, book} as a theory with such properties and also a realistic inflationary dynamics. Its distinctive feature is that diffeomorphism invariance in the ultraviolet (UV) limit is described as a special conformal invariance \cite{riegert, am, amm92, amm97, hh, hamada12M4, hamada12RS3}. Since it is a gauge symmetry, it represents background freedom in which all different conformally-flat spacetimes are gauge equivalent. This symmetry is called BRST\footnote{
BRST is an acronym consisting of the initials of Becchi, Rouet, Stora, and Tyutin, who discovered BRST symmetry of gauge theories in 1970s.} 
conformal invariance, which was first found in two-dimensional quantum gravity \cite{polyakov, kpz, dk, david} and then extended to four dimensions. The existence of this symmetry makes it possible to describe the world far beyond the Planck scale where scalar fluctuations dominate and no tensor ones.

Perturbation theory is formulated by introducing a dimensionless coupling constant $t$ that represents  deviation from the conformal invariance. The asymptotic background freedom represents that nonperturbative fluctuations of the conformal mode in the gravitational field become dominant in the UV limit of $t \to 0$. In this paper, we investigate how the universe evolves and deviates from such a scale-invariant state. The moment when the conformal invariance is completely broken at a novel energy scale of quantum gravity, $\Lam_\QG$, is called the ``spacetime phase transition". If we set this scale below the Planck scale, then the theory has an inflationary solution with the expansion time constant of order of the Planck mass \cite{hy, hhy06, hhy10}.

The goal of this study is to clarify how perturbations, or spacetime fluctuations, around the inflationary solution evolve, and to derive the primordial spectra that are  initial conditions of the current universe. It can be said that in the universe before inflation begins, quantum fluctuations of gravity are large and thus spacetime is not substantially fixed. As inflation begins, the fluctuations gradually decreases in amplitude,  eventually reaching the magnitude suggested by observations of tiny anisotropies in the cosmic microwave background radiation (CMB) \cite{wmap13, planck} until the time of the spacetime phase transition. Since fluctuations in gravity represent fluctuations in time and distance, the reduction of them implies that the universe becomes a real world where time and distance can be measured accurately.\footnote{
In the first place, there is no absolute time in gravitational theory, as we can see from the fact that the total Hamiltonian vanishes. Time is nothing but a dynamical variable that monotonically increases on average. Hence, the graviton picture propagating in a specific spacetime becomes applicable only after the spacetime phase transition when the fluctuations become sufficiently small.} 

In order to obtain the primordial spectrum, various nonlinear effects in the evolution of the fluctuations must be taken into account properly. The running coupling constant is one of these nonlinear effects, which is incorporated into the equations of motion by approximating it as a time-dependent mean field, as done in previous studies \cite{hhy06, hhy10}. In this way, we show that the amplitude of the fluctuations reduces during inflation.

To determine the spectral pattern more accurately, we need to evaluate contributions from nonlinear terms, such as the exponential factor of the conformal mode existing in the Einstein-Hilbert action. The fluctuations dealt with here are of sizes that momentum dependence disappears as soon as inflation begins, but the nonlinear effects cannot be ignored in the early stage where the fluctuations are still large.\footnote{
The strength of the nonlinearity is of the order of unity when expressed in terms of the non-Gaussianity parameter $f_{\rm NL}$ \cite{ks}.} 
One of the purposes of this research is to determine a system of nonlinear evolution equations of the fluctuations involving such effects.

\begin{figure}[h]
\begin{center}
\includegraphics[scale=0.32]{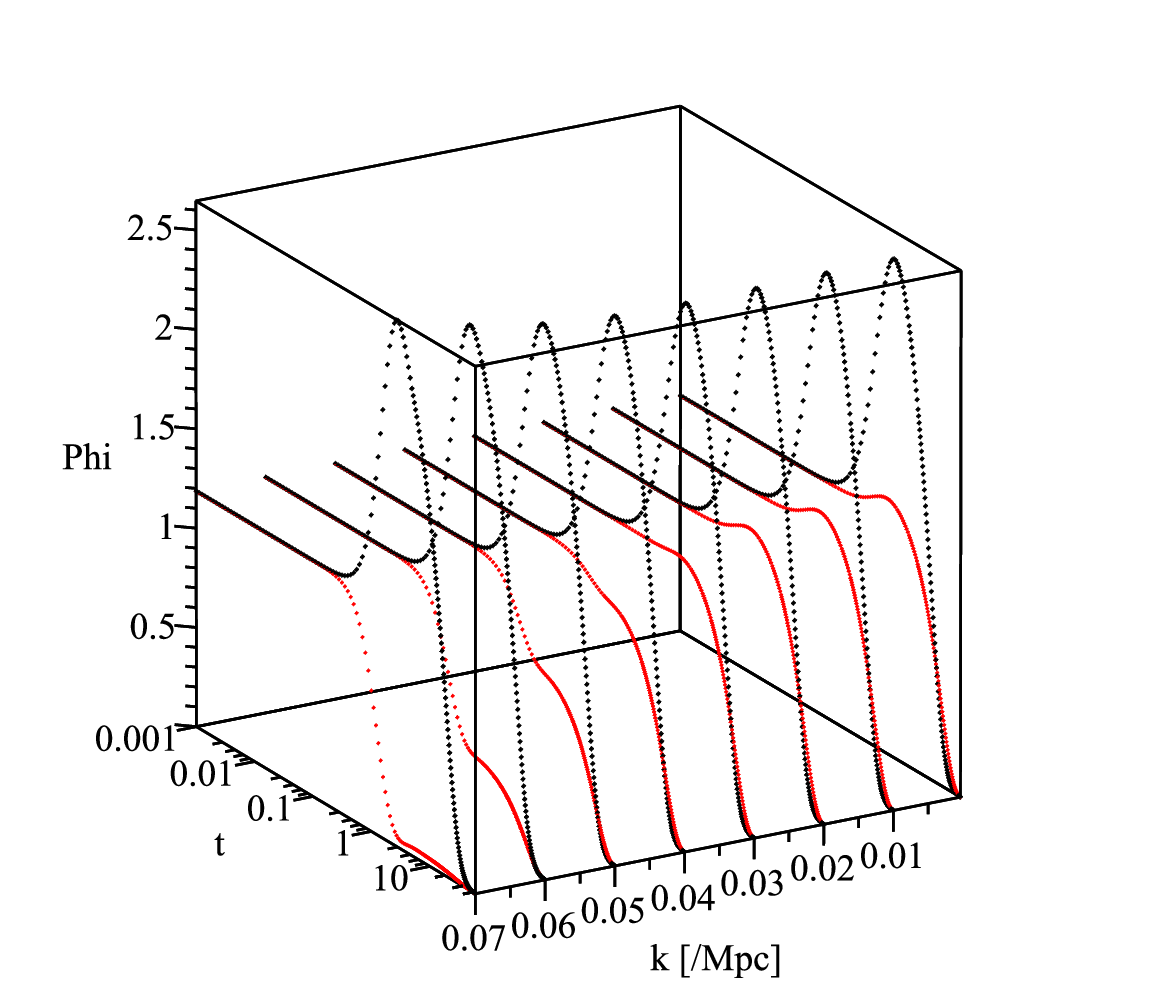}
\includegraphics[scale=0.32]{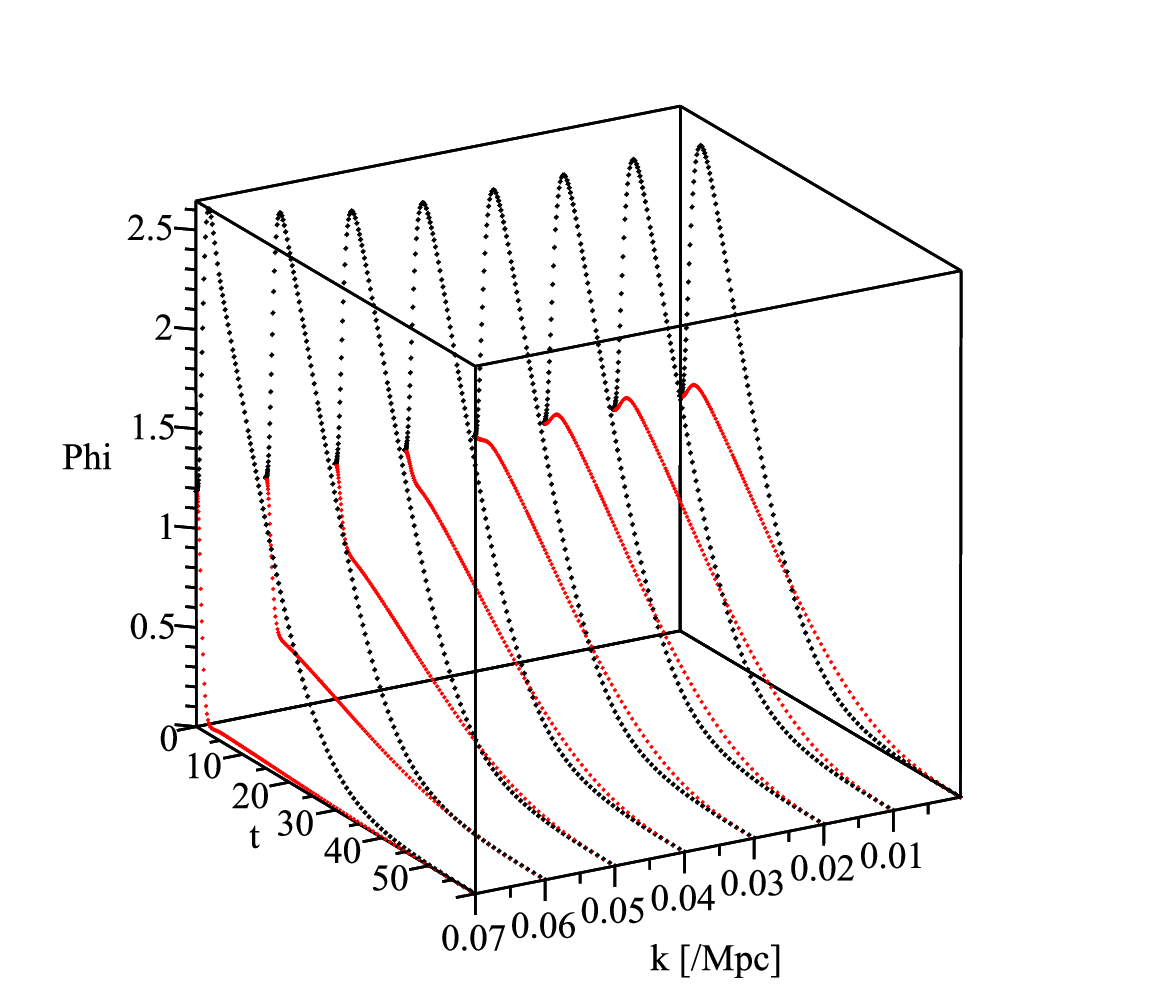}
\end{center}
\caption{\label{evolution of Phi in log and normal time}{\small 
The evolution of quantum gravity fluctuations incorporating nonlinear terms (black), together with previous linear equation results (red), from before inflation to the spacetime phase transition. The left is displayed in logarithmic time and the right is in normal time. The Planck time at which inflation begins is normalized to unity, and then the phase transition occurs at $60$. The horizontal axis $k$ is the comoving momentum.}}
\end{figure}

\begin{figure}[h]
\begin{center}
\includegraphics[scale=0.32]{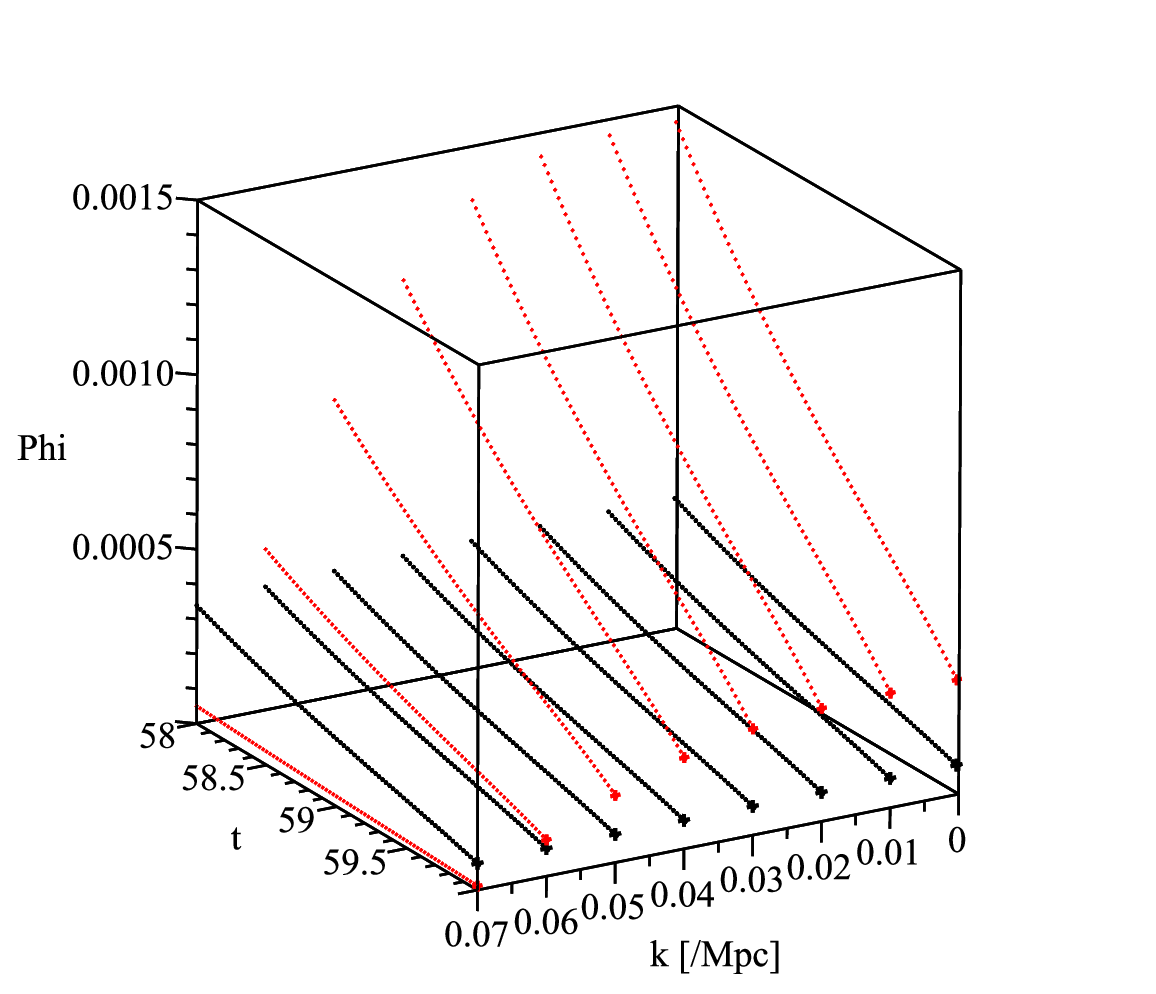}
\includegraphics[scale=0.26]{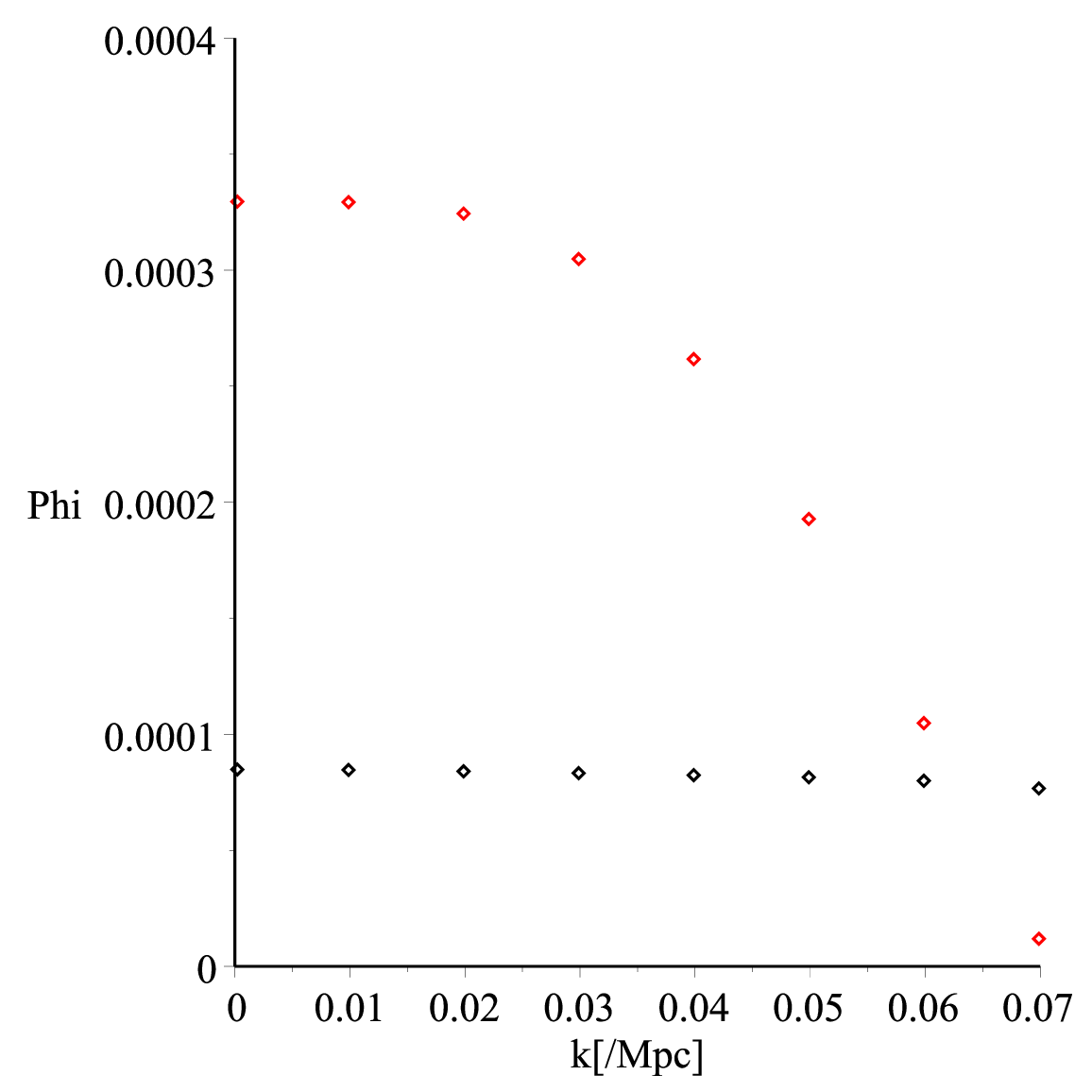}
\end{center}
\caption{\label{last stage and primordial spectrum}{\small The left is an enlarged view near the transition point in Fig.\ref{evolution of Phi in log and normal time}. The right is the values at the transition point, providing the primordial spectrum (a line connecting points, amplitude is its square).}}
\end{figure}

Since it is a rather complicated equation system, we here simplify it by extracting most contributing parts of the nonlinear terms in comoving momentum space so that we can solve it numerically and quantify their effects, as will be described in Sections 4 and 5. One of the results indicating that there is a large effect on the evolution of fluctuations is shown in Figs.\ref{evolution of Phi in log and normal time} and \ref{last stage and primordial spectrum}, together with previous results from linear equations. Consequently, we confirm that the nonlinear terms have expected effects by which the initial scale invariance is maintained up to relatively-high momentum regions, as inferred in the previous works \cite{hhy10}.

In determining the spectrum, the presence of the physical infrared cutoff $\Lam_\QG$ plays an essential role. The sharp falloff in low multipole components of the CMB angular power spectrum is explained by this scale \cite{hy}. Its wavelength is given by a comoving scale of the correlation length $\xi_\Lam =1/\Lam_\QG$. This means that most of the universe we see today was within the size of $\xi_\Lam$ before inflation.

Quantum gravity inflation suggests that the observed CMB anisotropy spectrum contains real quantum gravity effects. The guiding principles here are clear, diffeomorphism invariance and renormalizability. Thus the model gives life to the concept of inflation, aligning with Starobinsky's perspective \cite{starobinsky}. Moreover, we expect that the primordial spectrum derived from the first principle will settle the Hubble tension \cite{hubble-tension}.

\section{Renormalizable and Asymptotically Background-Free Quantum Gravity}
\setcounter{equation}{0}

\noindent
The problems with the Einstein's theory of gravity are summarized in unrenormalizability and the existence of singularities. Therefore, a finite UV cutoff is usually introduced in the Planck scale to avoid such problems. This corresponds to thinking of spacetime discretized in the Planck length, and some researchers regard the length as a quantum of spacetime. However, introducing the UV cutoff breaks diffeomorphism invariance.

If we consider diffeomorphism invariance as a first principle, then we have to bring the cutoff to infinity, i.e., take the continuum limit. It implies that the theory is renormalizable. Being renormalizable is equivalent to having no singularity. The reason why the existence of singularities cannot be denied probabilistically is due to  the fact that the Einstein-Hilbert action becomes finite for such singular spacetime configurations, because it does not contain the Riemann tensor $R_{\mu\nu\lam\s}$ which controls the magnitude of curvature. A renormalizable quantum theory of gravity requires an action involving the Riemann tensor squared with correct sign. Since such an action is positively divergent for singularities, they are eliminated as unphysical.\footnote{
This is a general argument without premising finiteness. When considering perturbation expansions around an apparently finite spacetime, the $R_{\mu\nu\lam\s}^2$ term is often removed using the Euler (Gauss-Bonnet) combination.} 
Furthermore, since the action becomes dimensionless, the corresponding coupling constant is also dimensionless and renormalizable \cite{stelle, tomboulis, ft}.

Despite these excellent properties, it is generally known that another problem arises in such a theory. It is the so-called ghost problem. However, note here that the existence of ghost modes itself is not a problem. In fact, Einstein's theory of gravity has a ghost mode due to the indefiniteness of the Einstein-Hilbert action. Owing to that, there are non-trivial solutions such as the Friedmann solution even though we are considering an equation that the total Hamiltonian vanishes. If the action was positive definite, there should only be a trivial vacuum solution. Thus, the ghost mode is an indispensable element to make the Hamiltonian zero, i.e., to preserve diffeomorphism invariance. However, it causes problems if it appears locally as a real object. A problem with early renormalizable theories using curvature squared is that the ghost mode appears as a physical particle state in the UV limit.

Considering why such a problem occurs, it turns out that there is a problem in the method of perturbation expansion. The conventional perturbation expansion has been carried out in the graviton picture propagating in flat spacetime of $R_{\mu\nu\lam\s}=0$, but this expansion method cannot make the ghost mode unphysical. 
To begin with, assuming that flat spacetime appears asymptotically implies that the closer we approach the center of the black hole, the weaker the gravitational field. Thus, this is a physically unnatural setting. Even theoretically, this perturbation method exhibits inconsistent UV behavior (see also footnote \ref{problem in conventional method}).

How should we perform perturbation expansion to correctly describe features of the world beyond the Planck scale? A solution to this problem can be found by considering what is required in the early universe. The important fact here is that inflation is given by a spacetime configuration in which the Weyl tensor
\begin{eqnarray*}
     C_{\mu\nu\lam\s} &=& R_{\mu\nu\lam\s} 
          -\half g_{\mu[\lam} R_{\s]\nu}  + \half g_{\nu[\lam} R_{\s]\mu} + \fr{1}{6} g_{\mu[\lam} g_{\s]\nu}  R
\end{eqnarray*}
vanishes, where $a_{[\mu} b_{\nu]} = a_\mu b_\nu - a_\nu b_\mu$. Therefore, we perform perturbation expansion around the conformally-flat spacetime with $C_{\mu\nu\lam\s}=0$.

This suggests that the conformal mode of the gravitational field, crucial for determining distances, receives special treatment. Here, we extract it in an exponential factor, ensuring its positivity
\begin{eqnarray*}
    g_{\mu\nu} = e^{2\phi} \bg_{\mu\nu} ,
\end{eqnarray*}
where the scalar-like field $\phi$ is called the conformal-factor field. The remaining mode $\bg_{\mu\nu}$ is expanded by the traceless tensor field $h_{\mu\nu}$ as
\begin{eqnarray}
   \bg_{\mu\nu} 
   = \eta_{\mu\lam} \, \bigl(e^{h} \bigr)^\lam_{~\nu}  = \eta_{\mu\lam} \biggl( \dl^\lam_{~\nu}+  h^\lam_{~\nu} 
                     + \fr{1}{2} h^\lam_{~\s} h^\s_{~\nu} + \cdots \biggr) ,
            \label{expansion of traceless tensor field}         
\end{eqnarray}
where $h^\mu_{~\mu}=0$. Raising and lowering the legs of the traceless tensor field is done with the flat background metric $\eta_{\mu\nu} =(-1,1,1,1)$ as $h_{\mu\nu}= \eta_{\mu\lam} h^\lam_{~\nu}$.
The coordinates of the flat metric are denoted as $x^\mu =(\eta, \bx)$, and are henceforth called the comoving coordinate, following cosmology.

We introduce a coupling constant $t$ as a dimensionless parameter that controls the above perturbation expansion. Noting that the field strength of the traceless tensor field is the Weyl tensor, the perturbation theory is defined by introducing the inverse of $t^2$ before the Weyl tensor squared with correct sign. The action of the whole system including it is given by \cite{hs, hamada02, hamada14re, hm16, book}
\begin{eqnarray*}
   I = \int d^4 x \hbox{$\sq{-g}$}  \biggl\{
      -\fr{1}{t^2} C_{\mu\nu\lam\s}^2 -b G_4 
       + \fr{1}{\hbar} \left( \fr{1}{16\pi G}R   + {\cal L}_\M \right) \biggr\} .
\end{eqnarray*}
The first term is called the Weyl action and is conformally invariant, $\sq{-g} \, C^2_{\mu\nu\lam\s}= \bar{C}^2_{\mu\nu\lam\s}$, where the quantity with the bar is defined by $\bg_{\mu\nu}$. The second $G_4=R_{\mu\nu\lam\s}R^{\mu\nu\lam\s}-4R_{\mu\nu} R^{\mu\nu} + R^2$ is the Euler density and its volume integral is also conformally invariant, where $b$ is not an independent coupling constant because this term does not contain a kinetic term, which is expanded in $t$. The third is the Einstein-Hilbert action and ${\cal L}_\M$ denotes the matter field actions which are conformally invariant in the UV limit. The cosmological constant term is ignored here because its contribution is negligible in the early universe. When performing perturbation calculations, we usually redefine the field as $h_{\mu\nu} \to th_{\mu\nu}$ in the expansion formula (\ref{expansion of traceless tensor field}), but here we proceed without redefining because we consider even regions where the coupling constant becomes large.

The coupling constant $t$ represents the degree of deviation from the conformal flatness. The significant point of this expansion method is that in the UV limit of $t \to 0$, the traceless tensor field becomes small, but the conformal-factor field $\phi$ remains fluctuating nonperturbatively. Thus, there is no picture of particles propagating in flat spacetime. The partition function under this expansion is given by \cite{hs}---\cite{david}
\begin{eqnarray*}
    \int [dg]_g \, e^{iI(g)} = \int [d\phi dh]_\eta \,e^{i S(\phi,\bg)+i I(g)} ,
\end{eqnarray*}
where the $\phi$-dependent action $S$ is the Wess-Zumino action \cite{wz} for conformal anomaly \cite{cd, ddi, duff, duff94}. Unlike the conventional perturbation expansion around flat spacetime, $S$ is necessary to preserve diffeomorphism invariance when replacing the path integral measure from the invariant $[dg]_g$ to the commonly used $[d\phi dh]_\eta$ defined on the flat background.\footnote{
\label{problem in conventional method}If we define the perturbation theory around flat spacetime, we need to introduce $R^2$ as a kinetic term for the conformal mode. However, the problem arises that a new coupling constant of this term does not exhibit asymptotically free behavior when it is introduced with correct sign \cite{ft}.} 
This is a quantity that makes up the running coupling constant introduced later, which represents the breaking of conformal invariance. Although the word anomaly is used, it is not physically anomalous. If formulated using dimensional regularization that preserve diffeomorphism invariance manifestly, the information of $S$ is automatically contained between $4$ and $D~(<4)$ dimensions \cite{hamada02, hamada14re, hm16, book, ft, hamada20CA, acd, bc, hathrellS, hathrellQED, hamada14GC}.

The Wess-Zumino actions are responsible for fourth-derivative dynamics of the conformal-factor field $\phi$. The most important term among them is the Riegert action \cite{riegert}, which remains even in $t \to 0$:\footnote{
The last $\bar{R}^2$ term is a loop correction required within the approximation adopted here, which compensates for quadratic terms of the traceless tensor field present in the nonlocal Riegert action \cite{riegert}.} 
\begin{eqnarray}
  S_{\rm R} =  \int d^4 x  \biggl\{ - \fr{b_1}{(4\pi)^2} \biggl( 2\phi \bDelta_4 \phi  + \bar{E}_4 \, \phi 
                         + \fr{1}{18}  \bar{R}^2 \biggr) \biggr\},
               \label{riegert action}          
\end{eqnarray}
where $E_4= G_4 -2\nb^2 R/3$ and $\nb^2=\nb^\mu \nb_\mu$. The differential operator $\sq{-g} \Delta_4$, which is conformally invariant for scalars, is defined by $\Delta_4 = \nb^4 + 2R^{\mu\nu} \nb_\mu \nb_\nu - 2R \nb^2/3 + \nb^\mu R \nb_\mu/3$. The coefficient is expanded as
\begin{eqnarray}
   b_1 = b_c B,  \qquad  B = 1 - \gm_1 \fr{t^2}{4\pi}  + o(t^4) .
         \label{expansion of coefficient b_1}
\end{eqnarray}
The lowest is given by $b_c = ( N_X + 11 N_W/2+ 62 N_A )/360 + 769/180$ with a right positive value, where
$N_X$, $N_W$, and $N_A$ are the numbers of scalar fields, Weyl fermions, and gauge fields coupled with gravity, respectively \cite{ft, amm92, hs}. For example, $b_c=7.0$ for the Standard Model, $b_c = 9.1$ for $SU(5) \,$GUT, and $b_c=12.0$ for $SO(10) \,$GUT. Here we adopt $b_c=7$.

Initial quantum gravity spectrum at $t=0$ is given by a two-point function of $\phi$. It is a logarithmic function of distance \cite{am, hamada12M4, hhy10}
\begin{eqnarray}
      \langle  \phi(\eta, \bx) \phi(\eta, \by) \rangle = - \fr{1}{4b_c} \log (\bx-\by)^2 .
              \label{two point function}
\end{eqnarray}
Fourier transforming this in three-dimensional space provides a scale-invariant scalar spectrum whose amplitude is a positive-definite constant proportional to the reciprocal of $b_c$. This fact is a consequence of the gravitational field being a dimensionless field.

The dynamics at the UV limit, where the Riegert action plays a central role, is described by the BRST conformal field theory, i.e., a special conformal field theory with conformal invariance as gauge symmetry.\footnote{
The transformation law for each field is expressed as $\dl_\zeta \phi = \zeta^\lam \pd_\lam \phi + \pd_\lam \zeta^\lam/4$ and $\dl_\zeta h_{\mu\nu} = \zeta^\lam \pd_\lam h_{\mu\nu} + ( \pd_\mu \zeta_\lam - \pd_\lam \zeta_\mu ) h^\lam_{~\nu}/2 + ( \pd_\nu \zeta_\lam - \pd_\lam \zeta_\nu ) h^\lam_{~\mu}/2$ with a gauge parameter $\zeta^\lam$ satisfying conformal Killing equations. The key here is that both right-hand sides are field-dependent. In contrast, in conventional weak-field approximation, gauge transformations remaining in the UV limit do not depend on fields, thus all modes that cannot be removed by gauge fixing become physical.} 
This is a hidden quantum field theory that emerges when rewriting quantum gravity into the theory on a particular background, here $\eta_{\mu\nu}$, and performing the path integral over $\phi$. Unlike normal conformal invariance, this invariance shows background freedom that expresses gauge equivalence between different conformally-flat spacetimes. Under this symmetry, all ghost modes involved in the $\phi$ and $h^\mu_{~\nu}$ fields become unphysical---not gauge invariant. Physical states are given by scalar composite states (primary scalar states) only, no tensor-type states. Thus, the asymptotic background freedom indicates that scalar fluctuations predominate in the early universe, consistent with observations.\footnote{
The BRST conformal invariance suggests that there are no tensor fluctuations of the CMB scales originated before inflation. Thus, they do not give a limit on the inflation scale so that we can go over the Planck scale wall.} 
The argument based on the conformal algebra can be found everywhere, which corresponds to solving the Hamiltonian and momentum constraints that bind ghosts \cite{am, amm92, amm97, hh, hamada12M4, hamada12RS3}.

As mentioned before, the unphysical ghost mode is a necessary ingredient for constructing a state in which the total Hamiltonian vanishes. If all modes were positive-definite, only such a state would be the trivial vacuum. Thus, the ghost mode in the gravitational field should be a hidden entity that does not appear locally, but necessary to preserve diffeomorphism invariance and to give entropy of the universe.\footnote{
The ghost is reminiscent of the Bohm's ``hidden variable'' in the sense that it is ``invisible, but being''. Reconciling gravity and quantum theory requires a deep understanding of what its being means. In relation to the Bohm interpretation of quantum cosmology, see for instance \cite{pinto-neto}, in which cosmological issues such as how to avoid singularities and achieve isotropizing are addressed.} 
This means that the quantum theory of gravity is not a theory for dealing with local entities such as gravitons, but a theory for describing state-changes in spacetime. Ghost modes do not appear directly in calculations that describe the evolution of quantum spacetime.  Although they definitely contribute to the evolution, they are not directly observed. This is true not only for quantum gravity but also for classical Einstein gravity. Evolution in the Friedmann universe is exactly due to the ghost mode, and the inflationary universe is as well.

Infrared dynamics of the traceless tensor field are represented by the running coupling constant. Defining the beta function as $\mu dt/d\mu = - \b_0 t^3 \, (\b_0 >0)$ with an arbitrary mass scale $\mu$ introduced upon quantization, it is given by\footnote{
The coefficient of the beta function is calculated as $\b_0 = [ ( N_X + 3 N_W+ 12 N_A )/240 + 197/60 ]/(4\pi)^2$ \cite{ft, amm92, hs}. However, here we will treat $\b_0$ as one of phenomenological parameters describing dynamics of strong coupling.} 
\begin{eqnarray}
     {\bar t}^2(Q) = \bigl[ \b_0 \log (Q^2/\Lam^2_{\rm QG}) \bigr]^{-1} , 
          \label{running coupling constant}    
\end{eqnarray}
where $Q^2 = q^2/e^{2\phi}$ and $q^2$ is the momentum squared defined on flat spacetime \cite{hamada02, hhy06, hamada20CA}. In cosmology, $Q$ is called the physical momentum and $q$ is called the comoving momentum. The energy scale $\Lam_\QG (=\mu e^{-1/2\b_0 t^2})$ is a renormalization group (RG) invariant satisfying $d \Lam_\QG/d\mu=0$, i.e., one of physical constants that do not change in the process of cosmic evolution \cite{collins}.

The effective action $\Gm$ including quantum corrections is given by replacing $t^2$ with ${\bar t}^2(Q)$. Specifically, the Weyl part of the effective action is $\Gm^{\rm W} = -[1/t^2 - 2\b_0 \phi + \b_0 \log (q^2/ \mu^2) ] \bar{C}^2_{\mu\nu\lam\s}$, where the second is one of the Wess-Zumino actions and the third is a loop correction, and then we can find that the inside of the square brackets can be summarized in the form of $1/{\bar t}^2 (Q)$ \cite{hamada02, hhy06, hamada20CA}. Thus, the $\phi$-dependence of the momentum $Q$ comes from the Wess-Zumino action $S$. At higher orders of $t$, multipoint interactions of the type $\phi^n \bar{C}_{\mu\nu\lam\s}^2$ arise, which are incorporated into $Q$ in higher-order corrections of the running coupling constant \cite{hamada20CA}. The Riegert part of the effective action, $\Gm^{\rm R}$, is also given by replacing the $t^2$-dependence in the coefficient (\ref{expansion of coefficient b_1}) with $\bar{t}^2(Q)$. In this case, the Wess-Zumino action of the type $\phi^{n+1} (2 \bDelta_4 \phi + \bar{E}_4)$ contributes. Similarly, in the presence of gauge field $F_{\mu\nu}$, the Wess-Zumino action of the type $\phi^n {\bar F}_{\mu\nu}^2$ arises and gauge coupling constant is then replaced with its running coupling constant \cite{hamada20CA}.

The effective action is a RG invariant, and the Planck mass and the cosmological constant appearing in it are also RG invariants, namely physical constants, such as the dynamical scale $\Lam_\QG$ \cite{hm17}. These three constants must be determined from observations. Here, the two mass scales $m_\pl$ and $\Lam_\QG$ are involved in the inflationary dynamics, while the cosmological constant is negligibly small. The magnitude relation must be $m_\pl > \Lam_\QG$ for inflation to occur. This relation implies that quantum gravity activates before reaching the Planck scale, i.e., the correlation length $\xi_\Lam$ becomes larger than the Planck length. The length $\xi_\Lam$ gives the size of localized quantum gravity excitations, leading to a dynamical discretization of spacetime \cite{hamada20LME}.

The relation $m_\pl > \Lam_\QG$ also serves as a unitarity condition to ensure that ghost particles do not appear in the world after the spacetime phase transition. In quantum spacetime, all ghost modes are constrained by BRST conformal invariance, but this cannot be applied after the transition. Therefore, the transition must occur below the Planck scale so that ghost gravitons with mass about $m_\pl$ do not appear.

The energy-momentum tensor is a formula obtained by applying a variation on the gravitational field to the effective action. Therefore, it is also a RG invariant and a finite operator, namely a normal product \cite{acd, bc, hathrellS, hathrellQED, hamada14GC}. The equation of motion for quantum gravity is expressed as the vanishing energy-momentum tensor. This is a general conclusion drawn from diffeomorphism invariance and renormalizability, implying no correction due to zero-point energy, which is consistent with no UV cutoff \cite{hamada22}. In addition, in the theory of gravity without absolute time, conservation of the energy-momentum tensor is guaranteed by the fact that it is a RG invariant.

Until now, we have paid no attention to $\hbar$, but here we clearly specify where it appears when restored. Considering the action $I$ by dividing it into the fourth-derivative gravitational part $I^{(4)}$ and the lower-derivative term $I_{\rm EG}$ consisting of the Einstein-Hilbert action and matter actions, the reciprocal of $\hbar$ is entered only in front of $I_{\rm EG}$. Since the gravitational field is essentially a dimensionless field, $I^{(4)}$ is completely dimensionless, thus it does not contain $\hbar$. Similarly, $\hbar$ is not  included in the Wess-Zumino action $S$ derived from the path integral measure. This means that all fourth-derivative gravitational actions are quantities that describe purely quantum dynamics, and the whole including them as weights, except $I_{\rm EG}$, could be regarded as a whole measure of the path integral. Hence, the path integral of quantum gravity is often symbolically expressed using only $I_{\rm EG}$ as $\int {\cal D}g_{\mu\nu} \, e^{iI_{\rm EG}/\hbar}$, but here it is shown that the measure is expressed exactly as ${\cal D} g_{\mu\nu} = [d\phi dh]_\eta \, e^{iS+iI^{(4)}}$. This is also the reason why the Weyl action and the Wess-Zumino action resulting from the measure are treated on the same footing in the following discussion.
From the consideration of $\hbar$, it turns out that the ghost mode is essentially a quantum entity and does not appear as a classical one like particles. In the following, $\hbar=1$.

\section{Inflationary Universe and Evolution Equation of Spacetime Fluctuations}
\setcounter{equation}{0}

\noindent
The idea of inflation has been conceived to solve the horizon problem of why there were larger correlations than the horizon size in the early universe and the flatness problem of why curvature remains near zero even after more than 10 billion years \cite{guth, sato, starobinsky}. Here we focus on another important role of inflation: generating initial conditions for evolution of the current universe, namely, the primordial spectra. They have to be almost scale-invariant and the magnitude of their amplitude has to be very small, at least for fluctuations with sizes involved in the structure formation of the universe. Inflation theory needs to describe these matters.

First, we show that the quantum gravity theory has an inflationary solution. The equation of motion that expresses evolution of the universe is derived by a variation of the effective action as follows \cite{hhy06}:
\begin{eqnarray}
    \dl \Gm = \int d^4 x \biggl( \bT^\mu_{~\mu} \dl \phi  +  \half \bT^\mu_{~\nu} \dl h^\nu_{~\mu} \biggl) = 0 .
         \label{definition of energy-momentum tensor}
\end{eqnarray}
The energy-momentum tensor is represented by the sum of the Riegert, Weyl, Einstein-Hilbert, and matter field sectors, as $\bT_{\mu\nu} = \bT^{\rm R}_{\mu\nu} + \bT^{\rm W}_{\mu\nu} + \bT^{\rm EH}_{\mu\nu} + \bT^{\rm M}_{\mu\nu}$, where $\bT_{\mu\nu} = \eta_{\mu\lam} \bT^\lam_{~\nu}$. Expressions for the energy-momentum tensors of each sector are summarized in Appendix A.

\subsection{Inflationary solution}

The conformal-factor field $\phi$ is divided into a spatially homogeneous component $\hphi$ and a fluctuation $\vphi$ around it as
\begin{eqnarray}
     \phi(\eta,\bx)=\hphi(\eta) +\vphi(\eta, \bx) .
          \label{fluctuation of phi}
\end{eqnarray}     
Homogeneous nonlinear equation for $\hphi$ is given from the trace equation $\bT^\mu_{~\mu}=0$ as \cite{hy, hhy06}
\begin{eqnarray}
       - \fr{b_c}{4\pi^2} B \pd_\eta^4 \hphi   
       + 6 M_\P^2 \, e^{2\hphi} \bigl(  \pd_\eta^2 \hphi    + \pd_\eta \hphi   \pd_\eta \hphi \bigr) =0 ,
             \label{homogeneous equation in conformal time}
\end{eqnarray}
where $M_\P=1/\sq{8\pi G}$ is the reduced Planck mass. The first term is from the Riegert action (\ref{riegert action}) and the second term is from the Einstein-Hilbert action, while there is no contribution from the Weyl and matter actions.  In addition, letting $\rho$ be an energy density of matter fields, we obtain an energy conservation equation from the time-time component equation $\bT_{00}=0$ as \cite{hhy06}
\begin{eqnarray}
     \fr{b_c}{8\pi^2} B \bigl( 2 \pd_\eta^3 \hphi \pd_\eta \hphi -\pd_\eta^2 \hphi \pd_\eta^2 \hphi  \bigr) 
      - 3 M_\P^2 \e^{2\phi} \pd_\eta \hphi \pd_\eta \hphi  + \e^{4\hphi} \rho = 0 .
         \label{energy conservation in conformal time}
\end{eqnarray}

Here we introduce the scale factor $a = e^\hphi$ and the proper time defined by $d\tau = a d\eta$. Further, introducing the Hubble variable $H=\pd_\tau a/a$ and their derivatives $Y=\pd_\tau H$ and $Z=\pd_\tau^2 H$, we can rewrite (\ref{homogeneous equation in conformal time}) as
\begin{eqnarray}
      B \bigl(  \pd_\tau Z   + 7 H Z     + 4 Y^2    + 18 H^2 Y     + 6 H^4  \bigr)  
      - 3 H_\D^2 \bigl(  Y + 2 H^2 \bigr) = 0 .
             \label{homogeneous equation in proper time}
\end{eqnarray}
When the coupling constant vanishes and thus $B = 1$, this equation has a stable inflationary solution with
\begin{eqnarray}
        H = H_\D,   \quad   H_\D=\sq{\fr{8\pi^2}{b_c}}M_\P .
            \label{inflationary solution}
\end{eqnarray}
The constant $H_\D$ has a value between the Planck mass $m_\pl$ and the reduced Planck mass $M_\P$ for the typical value of $b_c$ mentioned before. Therefore, it is also called the Planck scale. This solution expresses that the scale factor expands exponentially in the proper time as $a(\tau) \propto e^{H_\D \tau}$. Thus, the universe begins to grow rapidly from the Planck time $\tau_\P=1/H_\D$.

The energy conservation equation (\ref{energy conservation in conformal time}) can be rewritten as
\begin{eqnarray}
     B \bigl( 2H Z - Y^2  + 6H^2 Y +3 H^4 \bigr)    -3 H_\D^2 H^2   + \fr{8\pi^2}{b_c} \rho =0 .
          \label{energy conservation in proper time}
\end{eqnarray}
Substituting the solution (\ref{inflationary solution}) into this equation, we find that $\rho=0$ is an initial value.

\subsection{Spacetime phase transition}

As the space begins to expand, the running coupling constant $\bar{t}$ (\ref{running coupling constant}) increases accordingly. It implies that the universe gradually deviates from conformally-flat spacetime, and thus inflation terminates at the dynamical energy scale $\Lam_\QG$ where $\bar{t}^2$ diverges. At this time, the conformal gravity dynamics disappears completely and the classical spacetime phase emerges. In order to represent it as a time evolution, we approximate the running coupling constant by a time-dependent mean field \cite{hhy06, hhy10}. That is, replace $Q$ with the reciprocal of the proper time $\tau$ and express it as
\begin{eqnarray}
     \bar{t}^2(\tau) = \bigl[  \b_0 \log (\tau_\Lam^2/\tau^2) \bigr]^{-1} ,
        \label{time-dependent running coupling constant}
\end{eqnarray}
where $\tau_\Lam =1/\Lam_\QG$ is called the dynamical time. This running coupling constant increases logarithmically as inflation progresses, and diverges rapidly in the dynamical time. Then, the Weyl part of the effective action disappears in proportion to the reciprocal of $\bar{t}^2(\tau)$.

The effective action that describes infrared dynamics of the Riegert sector is also expressed by replacing $t^2$ in the factor $B \,$(\ref{expansion of coefficient b_1}) with $\bar{t}^2(\tau)$. We denote the factor that has undergone such a replacement as $\bar{B}(\tau)$ and consider the equation where $B$ in the equation of motion is simply replaced with $\bar{B}(\tau)$. Furthermore, in order to express disappearance of the dynamics of the Riegert sector with divergence of $\bar{t}^2(\tau)$, we assume the following form, called the dynamical factor \cite{hhy06, hhy10}:
\begin{eqnarray}
      \bar{B}(\tau) = \biggl[1+ \fr{\gm_1}{\kappa} \fr{\bar{t}^2(\tau)}{4\pi} \biggr]^{-\kappa} ,
           \label{time-dependent dynamical factor}
\end{eqnarray}
where $\gm_1, ~ \kappa >0$. The constants $\gm_1$, $\kappa$, and also $\b_0$ are here treated as phenomenological parameters that adjust the inflationary scenario. In this paper, these dynamical-factor parameters are set to be $\b_0=0.4$, $\gm_1=0.1$, and $\kappa=1$.

In this way, the spacetime phase transition is expressed as a process in which all the fourth-derivative gravitational actions responsible for the conformal gravity dynamics vanish at the dynamical time $\tau_\Lam$. The gravitational energy disappeared at that time is transferred to the energy of matter fields, causing the big bang. The energy conservation equation (\ref{energy conservation in proper time}) with $B$ replaced by $\bar{B}$ expresses such a process, that is, the matter energy $\rho$, which is initially zero, is generated when $\bar{B}$ disappears as the running coupling constant increases. The change occurs rapidly near the phase transition, and $\rho(\tau_\Lam)=3M_\P^2 H^2(\tau_\Lam)$ is obtained.

The number of e-foldings of inflation defined by ${\cal N}_e = \log [a(\tau_\Lam)/a(\tau_\P)]$ is roughly given by a ratio of the two physical scales $H_\D$ and $\Lam_\QG$. Here, we take the ratio to be
\begin{eqnarray}
      N = \fr{H_\D}{\Lam_\QG} = 60 .
          \label{ratio of two physical scales}
\end{eqnarray}    
In practice, the number of e-foldings ${\cal N}_e$ is slightly larger than this value, depending on the dynamical-factor parameters.

Since $b_c$ is set to $7$ here, the dynamical energy scale is determined from (\ref{ratio of two physical scales}) to be
\begin{eqnarray*}
      \Lam_\QG  = 1.3 \times 10^{17} \, {\rm GeV} 
\end{eqnarray*}
for $M_\P = 2.4 \times 10^{18} \,$GeV.

To clarify the physical meaning of the two mass scales, we introduce the corresponding comoving scales:
\begin{eqnarray}
      \lam = a(\tau_i) \Lam_\QG , \qquad  m =  a(\tau_i) H_\D, 
          \label{comoving energy scales}
\end{eqnarray}
where $a(\tau_i)$ is the scale factor at a time $\tau_i=1/E_i \, (E_i \! \gg \!H_\D)$ far before the Planck time, that is, before the scale factor starts to grow. The current scale factor is normalized to $1$ here.
Letting $\lam$ be a scale that explains the sharp falloff in low multipole components of the CMB angular power spectrum, its value is close to the Hubble constant $H_0$ and thus $a(\tau_i) \sim 10^{- 59}$ is derived. This implies that most of the universe we see today was originally created within the correlation length $\xi_\Lam$.

Furthermore, if the big bang occurred near $10^{17} \,$GeV and the present universe was created there, then the universe expanded about $10^{29}$ times after the big bang. From this, it is required to expand $10^{30}$ times during the inflationary period. Although $\Lam_\QG$ is not the only parameter determining the number of e-foldings, it turns out that this scale is a relatively tight one when considering the scenario of cosmic evolution. In this paper, we adopt $m=0.02 \, {\rm Mpc}^{-1}$ as a reference value, and then $\lam \,(=m/N) =0.00033 \, {\rm Mpc}^{-1}$.\footnote{
Using $1{\rm Mpc}=3.1 \times 10^{24} \, {\rm cm}$ and $1{\rm cm} = 5.1 \times 10^{13} \, {\rm GeV}^{-1}$, we obtain $a(\tau_i)=1.6 \times 10^{-59}$.} 

\begin{figure}[h]
\begin{center}
\includegraphics[scale=0.3]{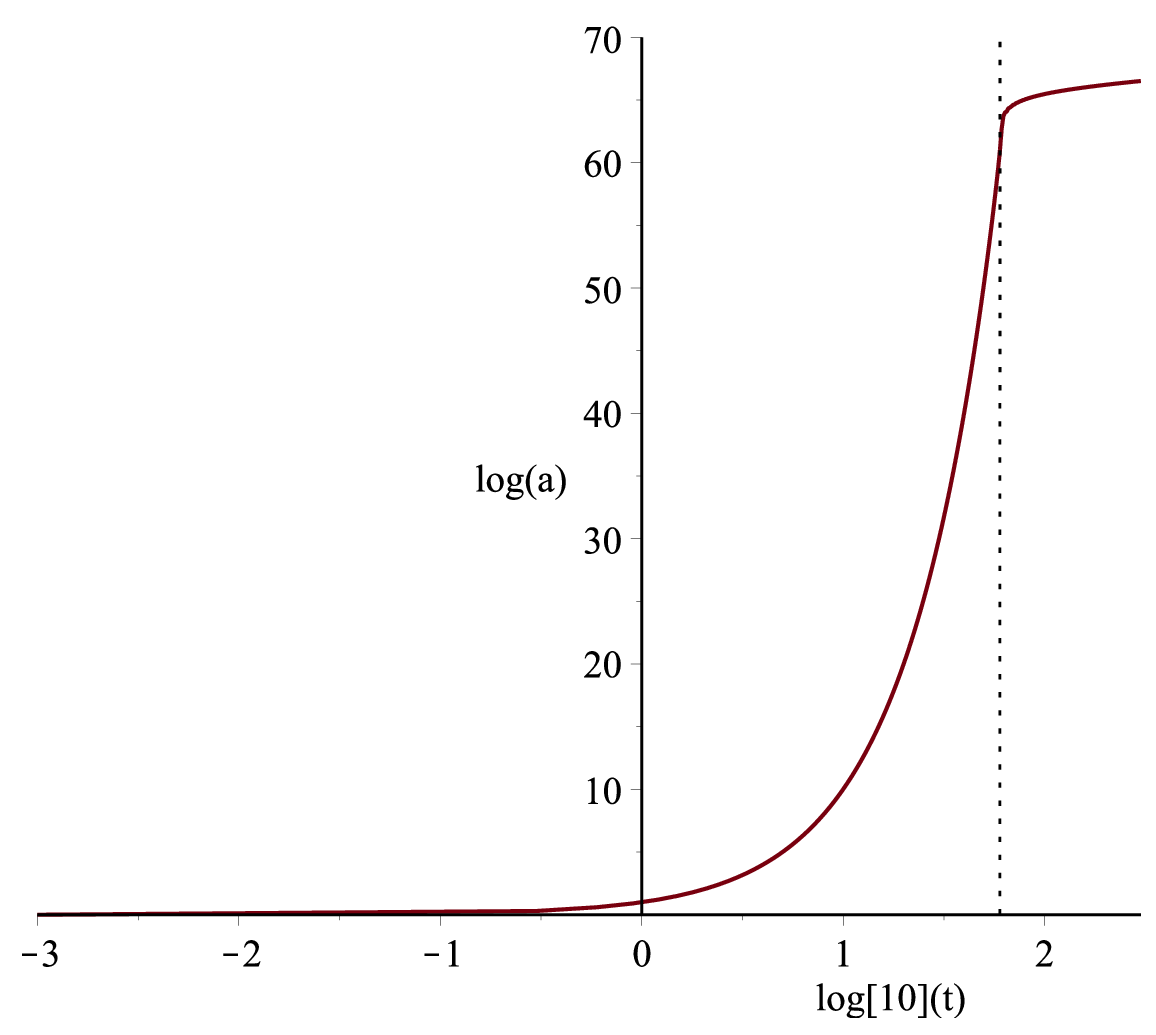}
\includegraphics[scale=0.3]{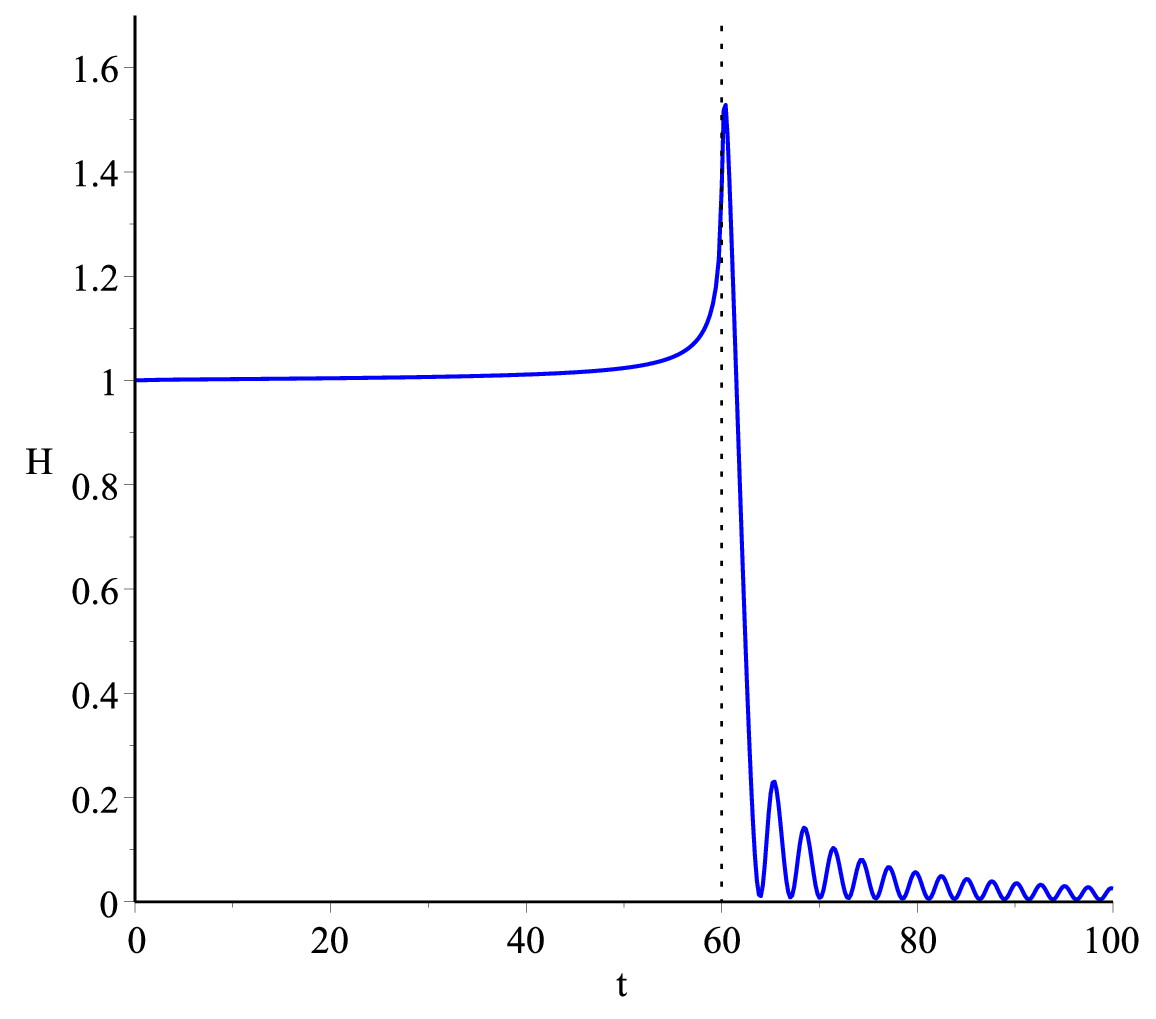}
\end{center}
\caption{\label{scale factor and H}{\small The left shows the time evolution of the scale factor $a$ and the right shows that of the Hubble variable $H$, where $H_\D$ is normalized to unity. The vertical dotted line represents the phase transition time $\tau_\Lam (= 60)$. The subsequent behavior is derived from a low-energy effective theory of gravity defined by derivative expansions in the gravitational field \cite{hhy06}.}}
\end{figure}

A numerical solution of the equation of motion for the homogeneous component is shown in Fig.\ref{scale factor and H}.\footnote{
At the phase transition point, the running coupling constant and the third derivative of $H$ diverges, but the physical quantities such as $\rho$ remains finite.}  
The horizontal axis is a dimensionless time defined by $t=H_\D\tau$ (not to be confused with the coupling constant). The Hubble variable $H$ is obtained by solving the homogeneous equation (\ref{homogeneous equation in proper time}) with $B$ replaced by $\bar{B}$. The matter energy density $\rho$ is then obtained by substituting the result of $H$ into (\ref{energy conservation in proper time}).
See Appendix B for details of the computation method.

We will now consider how the low-energy spacetime after the phase transition is described. In QCD, when the gluon action given by the second derivative disappears, the gluon dynamics disappears completely. Meanwhile, in quantum gravity, even if the fourth-derivative gravitational action disappears, the second-derivative Einstein-Hilbert action remains, which gives the kinetic term of the gravitational field at low energies. Thus, Einstein's theory of gravity appears as a low-energy effective theory. It is then no longer appropriate to consider separately the conformal-factor field and the traceless tensor field, and a composite field in which they are tightly coupled becomes a new variable of the gravitational field. The actual low-energy effective theory will be described by derivative expansions in the gravitational field with the Einstein-Hilbert action as the lowest order. According to this, the Hubble variable increases a little further after the phase transition, gains the e-foldings, and then oscillates and approaches the Friedmann solution rapidly, as shown in Fig.\ref{scale factor and H} (see \cite{hhy06} for details).

\subsection{Nonlinear evolution equation of fluctuations}

Here, we consider equations of motion for fluctuations around the inflationary solution. In fact, we should discuss the evolution of the physical scalars defined by composite fields, such as the quantum Ricci scalar. However, due to the fact that the scalar fluctuations reduce in amplitude during inflation, as previously mentioned, it turns out that it is sufficient to consider the evolution of the fluctuations, because the information we ultimately want to know is their reduced values at the phase transition, which gives the initial conditions of the Friedmann universe.

Now we derive a system of nonlinear evolution equations for the fluctuations. The conformal invariance indicates that the scalar fluctuation $\vphi \,$(\ref{fluctuation of phi}) dominates in the early universe beyond the Planck scale. In order to describe the process that the scalar fluctuation gradually deforms with time, we also take into account a scalar component of the traceless tensor field, defined by
\begin{eqnarray}
        h_{00} = h,  \qquad   h_{ij} = \fr{1}{3} \dl_{ij} h ,
            \label{h fluctuation}
\end{eqnarray}
which is initially vanishing.

The system of linear evolution equations for the fluctuations has already been derived, and it has been shown that the fluctuations gradually reduces during inflation \cite{hhy06, hhy10}. However, since $\vphi$ is still large in the initial stage, we cannot accurately estimate the fluctuation spectrum without incorporating nonlinear terms involving multipoint functions of $\vphi$ there. The exponential factor of $\vphi$ existing in the Einstein-Hilbert action provides such terms. On the other hand, $h$ is treated in the linear approximation, assuming that it is not very large, which is because $h$ is initially small. It then grows, but reduces again before the coupling constant $t$ becomes large. Thus, nonlinear terms are limited to first order in $h$.

In addition, scalar fluctuations contain nonlinear terms combining other tensor modes into scalars. These modes may also contribute to the equation of motion as nonlinear terms. However, since the initial smallness of these tensor modes is almost maintained \cite{hhy06}, the nonlinear terms containing them will contribute little to the evolution of the scalar fluctuations.

First, consider the trace part of the equations of motion. The Riegert sector is given by $\bT^{{\rm R}\mu}_{~~~\mu} = - (b_c/16\pi^2) B ( 4 \bDelta_4 \phi + \bE_4 )$. Expanding it to the first order of the traceless tensor field yields (\ref{trace of Riegert sector}). We expand this expression according to (\ref{fluctuation of phi}) and (\ref{h fluctuation}) and decompose it into the linear part of $\vphi$ and $h$ and the nonlinear part of $o(\vphi h)$. Writing it as $\bT^{{\rm R}\mu}_{~~~\mu} = \bT ^{{\rm R}\mu}_{~~~\mu} |_L + \bT^{{\rm R}\mu}_{~~~\mu} |_{NL}$, we obtain
\begin{eqnarray*}
     \bT^{{\rm R}\mu}_{~~~\mu} \vert_L 
     &=& \fr{b_c}{8\pi^2} B \biggl\{ 
              -2 \pd_\eta^4 \vphi      + 4 \pd_\eta^2 \lap3 \vphi     - 2 \dlap3 \vphi
              -4 \pd_\eta^4 \hphi h          -8\pd_\eta^3 \hphi \pd_\eta h   
              -\fr{16}{3}\pd_\eta^2 \hphi \pd_\eta^2 h
                   \nonumber \\
     &&
              + \fr{32}{9} \pd_\eta^2 \hphi \lap3 h      -\fr{4}{3} \pd_\eta \hphi \pd_\eta^3 h
              + \fr{20}{9} \pd_\eta \hphi \pd_\eta \lap3 h        - \fr{1}{3} \pd_\eta^4 h 
              + \fr{2}{9} \pd_\eta^2 \lap3 h       
              + \fr{1}{9} \dlap3 h    \biggr\}
\end{eqnarray*}
and
\begin{eqnarray*}
     \bT^{{\rm R}\mu}_{~~~\mu} \vert_{NL} 
     &=&  \fr{b_c}{8\pi^2} B \biggl\{  
             -4 h \pd_\eta^4 \vphi        + \fr{8}{3} h \pd_\eta^2 \lap3 \vphi 
             + \fr{4}{3} h \dlap3 \vphi      - 8 \pd_\eta h \pd_\eta^3 \vphi
             + \fr{8}{3} \pd^i h \pd_\eta^2 \pd_i \vphi    
                    \nonumber \\
     &&              
             + \fr{8}{3} \pd_\eta h \pd_\eta \lap3 \vphi    + \fr{8}{3} \pd^i h \pd_i \lap3 \vphi
             - \fr{16}{3} \pd_\eta^2 h \pd_\eta^2 \vphi    + \fr{32}{9} \lap3 h \pd_\eta^2 \vphi
             + \fr{16}{9} \lap3 h \lap3 \vphi          
                    \nonumber \\
     &&
             - \fr{4}{3} \pd_\eta^3 h \pd_\eta \vphi         + \fr{20}{9} \pd_\eta \lap3 h \pd_\eta \vphi    
             - \fr{4}{3} \pd_\eta^2 \pd^i h \pd_i \vphi       + \fr{4}{9} \pd^i \lap3 h \pd_i \vphi     \biggr\} .
\end{eqnarray*}
On the other hand, the trace part of the Weyl sector disappears in the first order of $h$, resulting in $\bT^{{\rm W}\lam}_{~~~~\lam}|_L=0$.

The trace part of the Einstein-Hilbert sector is given by $\bT^{{\rm EH}\mu}_{~~~~\mu} = M_\P^2 \sq{-g}R = M_\P^2 e^{2\phi}(\bR -6\bnb^2 \phi - 6 \bnb^\mu \phi \bnb_\mu \phi)$. Expanding it to the first order of the traceless tensor field, we get (\ref{trace of Einstein-Hilbert sector}). Expanding this further yields terms up to infinite order in $\vphi$ due to the presence of the exponential factor $e^{2\phi}$. Expressing it as $\bT^{{\rm EH}}_{\mu\nu} = \sum_{n=0}^\infty \bT^{\rm EH}_{(n) \mu\nu}$, $n=0$ is a homogeneous component that gives the second term of (\ref{homogeneous equation in conformal time}) and (\ref{energy conservation in conformal time}), and $\bT^{\rm EH}_{(n) \mu\nu}$ consists of the $o(\vphi^n)$ and $o(\vphi^{n-1} h)$ terms.  The linear part is given by
\begin{eqnarray*}
     \bT^{{\rm EH}\mu}_{(1) ~~\mu} 
         &=& M_\P^2 \, e^{2\hphi} \biggl\{
              6 \pd_\eta^2 \vphi          - 6 \lap3 \vphi      + 12 \pd_\eta \hphi \pd_\eta \vphi      
              + 12 \bigl( \pd_\eta^2 \hphi + \pd_\eta \hphi \pd_\eta \hphi  \bigr) \vphi
                     \nonumber \\
         &&
              + \pd_\eta^2 h         + \fr{1}{3} \lap3 h       + 6 \pd_\eta \hphi \pd_\eta h
              + 6 \bigl( \pd_\eta^2 \hphi + \pd_\eta \hphi \pd_\eta \hphi  \bigr) h   \biggr\}
\end{eqnarray*}
and the nonlinear part composed of $o(\vphi^2)$ and $o(\vphi h)$ is 
\begin{eqnarray*}
     \bT^{{\rm EH}\mu}_{(2)~~\mu} 
        &=& M_\P^2 \, e^{2\hphi} \biggl\{
             12 \vphi \pd_\eta^2 \vphi       - 12 \vphi \lap3 \vphi       + 6 \pd_\eta \vphi \pd_\eta \vphi
             - 6 \pd^i \vphi \pd_i \vphi      + 24 \pd_\eta \hphi  \, \vphi \pd_\eta \vphi
                  \nonumber \\
       &&
            + 12 \bigl( \pd_\eta^2 \hphi + \pd_\eta \hphi \pd_\eta \hphi  \bigr) \vphi^2        
            + 2 \vphi \pd_\eta^2 h      + \fr{2}{3} \vphi \lap3 h   
            + 6 \pd_\eta^2 \vphi h      + 2 \lap3 \vphi h     
                  \nonumber \\
       &&          
            + 6 \pd_\eta \vphi \pd_\eta h       + 2 \pd^i \vphi \pd_i h
            + 12 \pd_\eta \hphi \, \vphi \pd_\eta h    +  12 \pd_\eta \hphi \, \pd_\eta \vphi h
                    \nonumber \\
      &&
            + 12 \bigl( \pd_\eta^2 \hphi + \pd_\eta \hphi \pd_\eta \hphi  \bigr) \vphi h   \biggr\} .
\end{eqnarray*}
The case of $n \geq 3$ can also be easily derived.

Since there are two field variables, the trace equation alone is incomplete. Therefore, we consider another equation that combines the energy-momentum tensor such as
\begin{eqnarray}
     \fr{1}{\lap3} \biggl( \bT^i_{~i} -3 \fr{\pd^i \pd^j}{\lap3} \bT_{ij} \biggr) = 0 .
          \label{definition of constraint equation}
\end{eqnarray}
Unlike the trace equation, there is a contribution from the Weyl sector, which is derived from the linear part of the energy-momentum tensor given in (\ref{Weyl energy-momentum tensor}). 
On the other hand, there is no contribution from the matter sector in this case either.

Since there is $1/t^2$ in front of the Weyl part, this part dominates at the early stage when $t^2$ is small. Multiplying the whole equation of motion by $t^2$ to remove it, we can then see that the contribution from the Riegert and Einstein-Hilbert sectors becomes $o(t^2)$ so that they can be treated in linear in the early stage. The contribution to (\ref{definition of constraint equation}) from the Riegert sector is thus given by linear part of $\vphi$ and $h$ obtained by substituting (\ref{fluctuation of phi}) into (\ref{Riegert energy-momentum tensor}). The contribution from the Einstein-Hilbert sector is also derived from (\ref{Einstein-Hilbert energy-momentum tensor}).

Furthermore, since the fluctuations decrease as the coupling constant $t$ increases, Eq.(\ref{definition of constraint equation}) can remain the linear approximation from beginning to end. This equation is expressed by the second derivative of the fields as
\begin{eqnarray}
    &&  \fr{16}{3} \fr{1}{t^2} \biggl( \pd_\eta^2 h    - \fr{1}{3} \lap3 h \biggr)
          + \fr{b_c}{8\pi^2} B \biggl\{ \fr{4}{3} \pd_\eta^2 \vphi     + \fr{8}{3} \pd_\eta \hphi \, \pd_\eta \vphi
               +\biggl( 8 \pd_\eta^2 \hphi         - \fr{4}{3} \lap3  \biggr) \vphi 
                          \nonumber \\
    &&  + \fr{2}{9}  \pd_\eta^2 h      + \fr{4}{3} \pd_\eta \hphi \,  \pd_\eta h
          + \biggl( \fr{20}{9} \pd_\eta^2 \hphi        - \fr{16}{9} \pd_\eta \hphi \pd_\eta \hphi 
                       + \fr{2}{27} \lap3   \biggr) h       \biggr\}
                       \nonumber \\
    &&   + M_\P^2 \, e^{2\hphi} \biggl( - 4 \vphi     + \fr{2}{3} h \biggr)    = 0 .
             \label{constraint equation}                 
\end{eqnarray}
Thus, the system of evolution equations can be expressed as a coupled differential equation of this linear equation and the nonlinear trace equation $\bT^{{\rm R}\mu}_{~~~\mu} \vert_L + \bT^{{\rm R}\mu}_{ ~~~\mu} \vert_{NL} + \sum_{n=1}^\infty \bT^{{\rm EH}\mu}_{(n)~~\mu} = 0$.

However, it is not yet in a form that can be solved numerically. It is necessary to rewrite the $\pd_\eta^4 \hphi \, h$ term in the $\bT^{{\rm R}\mu}_{~~~\mu} \vert_L$ part to terms with lower time-derivatives of $\hphi$ using the homogeneous equation (\ref{homogeneous equation in conformal time}). In addition, noting that the trace equation has the structure
\begin{eqnarray*}
     && - \fr{b_c}{4\pi^2} B \bigl( \pd_\eta^4 \vphi -2 \pd_\eta^2 \lap3 \vphi + \dlap3 \vphi \bigr)
         + M_\P^2 \, e^{2\hphi} \bigl\{ 6 \pd_\eta^2 \vphi - 6\lap3 \vphi 
                \nonumber \\
     && 
              +12 \pd_\eta \hphi  \, \pd_\eta \vphi 
              + 12 \bigl( \pd_\eta^2 \hphi + \pd_\eta \hphi \pd_\eta \hphi \bigr) \vphi 
              \bigr\} = o(h) ,
\end{eqnarray*}
the $h \, \pd_\eta^4 \vphi$ term in the $\bT^{{\rm R}\mu}_{~~~\mu} \vert_{NL}$ part is rewritten to terms with lower time-derivatives of $\vphi$ by ignoring the $o(h^2)$ term.
In addition, we introduce a field variable
\begin{eqnarray*}
      \Phi=\vphi+\fr{1}{6} h
\end{eqnarray*}    
and treat $\Phi$ and $h$ as independent variables.  The fourth time-derivative of $h$ in the linear part is then absorbed into a fourth time-derivative of $\Phi$ so that the time-derivative of $h$ is given by up to at most the third one. Here, $\Phi$ is the gravitational potential when the line-element is expressed within linear as $ds^2 = a^2 [-(1+2 \Psi )d\eta^2 + (1+2 \Phi ) d\bx^2 ]$, where the other one is given by $\Psi=\vphi-h/2$.

The nonlinear terms that become effective in the early stage of the inflationary universe can be incorporated in this way. On the other hand, the nonlinear effect arising in infrared conformal gravity dynamics is represented by the running coupling constant. As mentioned earlier, it is expressed by adopting a type of the mean-field approximation that rewrites the coupling constant squared to the time-dependent average $\bar{t}^2(\tau) \,$(\ref{time-dependent running coupling constant}) and also $B$ to the time-dependent dynamical factor $\bar{B}( \tau) \,$(\ref{time-dependent dynamical factor}). While this seems like a radical model, no detailed information of strong coupling dynamics other than the disappearance of conformal gravity dynamics will be needed, because the size of the fluctuations considered here becomes much larger than the correlation length $\xi_\Lam$ near the phase transition point where the coupling constant increases.

The system of nonlinear evolution equations derived finally is given by
\begin{eqnarray}
       && \fr{b_c}{8\pi^2} {\bar B}(\tau) \biggl\{ 
             -2 \pd_\eta^4 \Phi    + 4 \pd_\eta^2 \lap3 \Phi     -2 \dlap3 \Phi 
             - \fr{4}{9} \pd_\eta^2 \lap3 h       +  \fr{4}{9}  \dlap3 h 
             - 8 \pd_\eta^3 \hphi  \, \pd_\eta h        
                        \nonumber \\
       &&
            - \fr{16}{3} \pd_\eta^2 \hphi  \, \pd_\eta^2 h       - \fr{4}{3} \pd_\eta \hphi  \, \pd_\eta^3 h       
            + \fr{32}{9}  \pd_\eta^2 \hphi \lap3 h           + \fr{20}{9} \pd_\eta \hphi  \, \pd_\eta \lap3 h
                       \nonumber \\
      &&
             - \fr{16}{3} h \pd_\eta^2 \lap3 \Phi        + \fr{16}{3} h \dlap3 \Phi   
             - 8 \pd_\eta h \pd_\eta^3 \Phi            + \fr{8}{3} \pd^i h \pd_\eta^2 \pd_i \Phi
             + \fr{8}{3} \pd_\eta h \pd_\eta \lap3 \Phi                
                       \nonumber \\
      &&
            + \fr{8}{3} \pd^i h \pd_i \lap3 \Phi        - \fr{16}{3} \pd_\eta^2 h \pd_\eta^2 \Phi
            + \fr{32}{9} \lap3 h \pd_\eta^2 \Phi      + \fr{16}{9} \lap3 h \lap3 \Phi      
            - \fr{4}{3} \pd_\eta^3 h \pd_\eta \Phi
                       \nonumber \\
      &&
            + \fr{20}{9} \pd_\eta \lap3 h \pd_\eta \Phi      - \fr{4}{3} \pd_\eta^2 \pd^i h \pd_i \Phi
            + \fr{4}{9} \pd^i \lap3 h \pd_i \Phi    \biggr\}   
                      \nonumber \\
      &&
          + M_\P^2 \, e^{2\hphi} \biggl\{ 
              6 \pd_\eta^2 \Phi           + 12 \pd_\eta \hphi  \, \pd_\eta \Phi       - 6 \lap3 \Phi
              + 12 \bigl( \pd_\eta^2 \hphi  + \pd_\eta \hphi \pd_\eta \hphi \bigr) \Phi
                      \nonumber \\
     &&
             + 4 \pd_\eta \hphi  \, \pd_\eta h          + \fr{4}{3} \lap3 h
             - 8 \bigl( \pd_\eta^2 \hphi  + \pd_\eta \hphi \pd_\eta \hphi \bigr) h 
                      \nonumber \\
     &&
             + 12 \Phi \pd_\eta^2 \Phi        + 6 \pd_\eta \Phi \pd_\eta \Phi    
             + 24 \pd_\eta \hphi  \, \Phi \pd_\eta  \Phi        
             + 12 \bigl( \pd_\eta^2 \hphi  + \pd_\eta \hphi \pd_\eta \hphi \bigr)  \Phi^2
                       \nonumber \\
     &&
             - 12 \Phi \lap3 \Phi           - 6 \pd^i \Phi \pd_i \Phi       
             - 8 h \pd_\eta^2 \Phi         + 4 \pd_\eta h  \pd_\eta \Phi 
             -16 \pd_\eta \hphi  \, h \pd_\eta \Phi         
             + 8 \pd_\eta \hphi  \, \pd_\eta h \Phi         
                      \nonumber \\
     &&
             - 16 \bigl( \pd_\eta^2 \hphi  + \pd_\eta \hphi \pd_\eta \hphi \bigr) h \Phi      
             + 16 h \lap3 \Phi             + 4 \pd^i h \pd_i \Phi
             + \fr{8}{3} \lap3 h \Phi     \biggr\}      = 0   
               \label{final trace equation}
\end{eqnarray}
and 
\begin{eqnarray}
     &&  \fr{16}{3} \fr{1}{\bar{t}^2(\tau)} \biggl(  \pd_\eta^2 h    - \fr{1}{3} \lap3 h  \biggr)
            + \fr{b_c}{8\pi^2} \bar{B}(\tau) \biggl\{  \fr{4}{3} \pd_\eta^2 \Phi     - \fr{4}{3} \lap3 \Phi   
                      +  \fr{8}{3} \pd_\eta \hphi  \, \pd_\eta \Phi      + 8 \pd_\eta^2 \hphi \, \Phi  
                           \nonumber \\
     &&  + \fr{8}{27} \lap3 h       + \fr{8}{9}  \pd_\eta \hphi  \, \pd_\eta h     
           + \biggl( \fr{8}{9} \pd_\eta^2 \hphi   - \fr{16}{9} \pd_\eta \hphi \pd_\eta \hphi \biggr) h  \biggr\}                      
           + M_\P \, e^{2\hphi}  \biggl(  - 4 \Phi    + \fr{4}{3} h \biggr) 
                       \nonumber \\
     && = 0  .               
             \label{final constraint equation}
\end{eqnarray}

The second equation (\ref{final constraint equation}) plays an important role as a constraint for connecting between the inflation and Friedmann phases. When the running coupling constant $\bar{t}$ is small, $h$ is small, indicating that $\vphi$ is dominant. Conversely, when $\bar{t}$ diverges, $1/\bar{t}^2$ and the dynamical factor $\bar{B}$ vanish, and the last Einstein term becomes dominant, this results in a transition to the Friedmann universe satisfying $\Phi=h/3$. In terms of the gravitational potentials, it expresses that the initial fluctuation of $\Phi=\Psi$ changes to $\Phi=-\Psi$.

The equations (\ref{final trace equation}) and (\ref{final constraint equation}) will be solved simultaneously. At that time, it is necessary that $\pd_\eta^3 h$ in (\ref{final trace equation}) is rewritten to the third time-derivative of $\Phi$ and other lower time-derivative terms using an equation obtained by time differentiating  (\ref{final constraint equation}) as
\begin{eqnarray}
   \pd_\eta^3 h &=& \fr{1}{3} \pd_\eta \lap3 h  
       + \fr{1}{{\bar t}^2(\tau)} \pd_\eta {\bar t}^2(\tau) \biggl( \pd_\eta^2 h  - \fr{1}{3} \lap3 h \biggr)
                        \nonumber \\
          &&  - \fr{b_c}{32\pi^2} {\bar t}^2(\tau) \biggl\{  {\bar B}(\tau) \biggl[ 
                   \pd_\eta^3 \Phi  + 2 \pd_\eta \hphi \pd_\eta^2 \Phi
                   + 8 \pd_\eta^2 \hphi \pd_\eta \Phi     + 6 \pd_\eta^3 \hphi  \Phi   
                   - \pd_\eta \lap3 \Phi 
                             \nonumber \\
          &&\quad
                  + \fr{2}{3} \pd_\eta \hphi \pd_\eta^2 h  
                  + \fr{4}{3} \bigl( \pd_\eta^2 \hphi  - \pd_\eta \hphi \pd_\eta \hphi \bigr) \pd_\eta h
                  + \fr{2}{3} \bigl( \pd_\eta^3 \hphi   - 4 \pd_\eta^2 \hphi \pd_\eta \hphi \bigr) h  
                  + \fr{2}{9} \pd_\eta \lap3 h  \biggr]           
                          \nonumber \\
         &&\quad
                + \pd_\eta {\bar B}(\tau) \biggl[  
                   \pd_\eta^2 \Phi  + 2 \pd_\eta \hphi \pd_\eta \Phi     + 6 \pd_\eta^2 \hphi  \Phi
                   - \lap3 \Phi     + \fr{2}{3}  \pd_\eta \hphi \pd_\eta h  
                              \nonumber \\
        &&\quad
                   + \fr{2}{3} \bigl( \pd_\eta^2 \hphi  - 2 \pd_\eta \hphi  \pd_\eta \hphi \bigr) h
                   + \fr{2}{9} \lap3 h  \biggr] \biggr\}
                              \nonumber \\
       && + M_\P^2 \, {\bar t}^2(\tau) e^{2\hphi} \biggl\{
                     \fr{3}{4} \pd_\eta \Phi   +  \fr{3}{2} \pd_\eta \hphi  \Phi
                     - \fr{1}{4} \pd_\eta h   - \fr{1}{2} \pd_\eta \hphi h   \biggr\}   .   
               \label{third derivative of h}                 
\end{eqnarray}

\section{How To Handle Nonlinear Terms}
\setcounter{equation}{0}

\noindent
The fluctuation spectrum is expressed using Fourier transform in the three-dimensional comoving coordinate space. Letting $f(\bx)$ be a dimensionless field variable, here we define its Fourier transform so that the transformed quantity is also dimensionless as
\begin{eqnarray*}
    f(\bx) = \int \fr{d^3 \bk}{(2\pi)^3}\fr{1}{k^3} f(\bk) \, e^{i\bk \cdot \bx} ,
\end{eqnarray*}
where $k =|\bk|$. Its dimensionless power spectrum $\langle |f(\bk)|^2 \rangle/2\pi^2$ is defined by $\langle f(\bk)f(\bk^\pp) \rangle = \langle |f(\bk)|^2 \rangle k^3 (2\pi)^3 \dl^3 (\bk+\bk^\pp)$. The scale invariance is expressed as this power spectrum being constant.

Fourier transform of a dimensionless field product $f(\bx)g(\bx)$ is also defined by
\begin{eqnarray}
     (fg)(\bk) &=& k^3 \int d^3 \bx f(\bx)g(\bx) e^{-i\bk \cdot \bx}
        \nonumber \\
        &=& k^3 \int d^3 \bx
            \int \fr{d^3 \bp}{(2\pi)^3} \fr{d^3 \bq}{(2\pi)^3} \fr{1}{p^3 q^3}
            f(\bp) g(\bq) e^{i(\bp + \bq - \bk)\cdot \bx} .
                \label{definition of fourier product}
\end{eqnarray}
Here, we rewrite the exponential part using the expansion formula with the spherical Bessel function,
\begin{eqnarray*}
    e^{i\bk \cdot \bx} = 4\pi \sum_{l=0}^\infty \sum_{m=-l}^l 
            i^l j_l (kx) Y_{lm}^*(\Om_k) Y_{lm}(\Om_x) ,
\end{eqnarray*}
where $\Om$ denotes angular components such that $d\bx^3 = x^2 dx d\Om_x$ and the spherical harmonics are normalized as $\int d\Om Y^*_{lm}(\Om) Y_{l^\pp m^\pp}(\Om) = \dl_{l l^\pp} \dl_{m m^\pp}$.

Henceforth, assuming that isotropic components of each function in the linear and nonlinear terms contribute predominantly, let 
\begin{eqnarray}
     f(\bp)=f(p),  \qquad g(\bq)=g(q) .
        \label{isotropic condition}
\end{eqnarray}
In this case, the angular integration in (\ref{definition of fourier product}) can be easily performed and we obtain
\begin{eqnarray*}
    (fg)(k) = \fr{8}{(2\pi)^3} k^3 \int \fr{dp}{p} \fr{dq}{q} 
               f(p) g(q) \int dx x^2 j_0(kx) j_0 (p x) j_0(q x) ,
\end{eqnarray*}
where $j_0(kx)=\sin (kx)/kx$. The $x$-integral yields
\begin{eqnarray*}
    \int_0^\infty d x x^2 j_0(kx)j_0(p x)j_0(q x)
    = \left\{ \begin{array}{ll}
               \fr{\pi}{4}\fr{1}{k p q}  & \mbox{for $|p - q|< k < p + q$}  \\
               0 & \mbox{otherwise} 
              \end{array}
              \right.
\end{eqnarray*}
and thus we get
\begin{eqnarray}
    (fg)(k) = \fr{k^2}{(2\pi)^2} \int \fr{dp}{p^2} \fr{dq}{q^2}  
                  \theta(k-|p-q|)\theta(p+q-k) f(p) g(q) ,
            \label{fourier transform of product fg}      
\end{eqnarray}
where $\theta$ is the Heaviside step function.

Fourier transform of a product of functions with spatial derivatives can be calculated similarly. Here it suffices for us to consider the following type:  
\begin{eqnarray*}
     && (\pd_if \pd^i g)(k) = k^3 \int d^3 \bx 
            \pd_i f(\bx) \pd^i g(\bx) e^{-i\bk \cdot \bx}
                \nonumber \\
      && \qquad
         = k^3 \int d^3 \bx 
            \int \fr{d^3 \bp}{(2\pi)^3} \fr{d^3 \bq}{(2\pi)^3}
             \fr{- \bp \cdot \bq}{p^3 q^3} f(\bp) g(\bq) 
             e^{i(\bp+\bq-\bk)\cdot \bx} .
\end{eqnarray*}
Under the isotropic condition (\ref{isotropic condition}), integrating the angular components of the momentum $\bp$ and $\bq$ can be easily performed by expressing the vector $\bp=(p_x,p_y,p_z)$ in terms of spherical harmonics as
\begin{eqnarray*}
     p_x &=& \sq{\fr{2\pi}{3}} p \left\{ Y_{1-1}(\Om_p)-Y_{11}(\Om_p) \right\},
                \nonumber \\
     p_y &=& i \sq{\fr{2\pi}{3}} p \left\{ Y_{1-1}(\Om_p)+Y_{11}(\Om_p) \right\},
                \nonumber \\
     p_z &=& \sq{\fr{4\pi}{3}} p Y_{10}(\Om_p) .
\end{eqnarray*}
Performing the integration of the remaining angular component $\Om_x$ using
\begin{eqnarray*}
    \int d\Om_x Y^*_{l_1 m_1}(\Om_x) Y_{l_2 m_2}(\Om_x) Y_{l_3 m_3}(\Om_x) 
    = \sq{\fr{(2l_2 +1)(2l_3 +1)}{4\pi(2l_1 +1)}} 
        C^{l_1 0}_{l_2 0 l_3 0} C^{l_1 m_1}_{l_2 m_2 l_3 m_3} 
\end{eqnarray*}
yields
\begin{eqnarray}
    (\pd_i f \pd^i g)(k)  &=& \fr{8}{(2\pi)^3} k^3 \int dp dq f(p) g(q) 
             \int dx x^2 j_0(kx) j_1 (p x) j_1(q x)
            \nonumber \\
     &=& \fr{k^2}{(2\pi)^2} \int \fr{dp}{p^2} \fr{dq}{q^2} 
                  \theta(k-|p-q|)\theta(p+q-k) 
            \nonumber \\
    && \qquad\qquad
         \times 
         \half (p^2 + q^2 - k^2) f(p) g(q) .
            \label{fourier transform of product dfdg}      
\end{eqnarray}

The lower bound of the comoving momentum integral is given by the physical comoving scale $\lam \,$(\ref{comoving energy scales}), while there is no upper bound on the integration. Using the formulas (\ref{fourier transform of product fg}) and (\ref{fourier transform of product dfdg}), Fourier transforms of the second-order nonlinear terms in the trace equation (\ref{final trace equation}) are given by
\begin{eqnarray}
    &&\fr{b_c}{8\pi^2} \bar{B}  \fr{k^2}{(2\pi)^2} \int^\infty_\lambda 
         \fr{dp}{p^2} \fr{dq}{q^2} \theta(k-|p-q|)\theta(p+q-k) 
          \biggl\{  - 8 \pd_\eta h(p) \pd_\eta^3 \Phi(q)          
                 \nonumber \\
    &&
          - \fr{16}{3} \pd_\eta^2 h(p) \pd_\eta^2 \Phi(q)
          - \fr{4}{3} \pd_\eta^3 h(p) \pd_\eta \Phi(q)  
           + \biggl( - \fr{20}{9} p^2   + \fr{20}{3} q^2   - \fr{4}{3} k^2 \biggr) h(p) \pd_\eta^2 \Phi(q)
                 \nonumber \\  
    &&
           + \biggl( - \fr{20}{9} p^2   -  \fr{8}{3} q^2 \biggr) \pd_\eta h(p) \pd_\eta \Phi(q)       
           + \biggl(  - \fr{2}{3} p^2    - \fr{2}{3} q^2   + \fr{2}{3} k^2 \biggr) \pd_\eta^2 h(p) \Phi(q)    
                 \nonumber \\
    &&
           + \biggl[  - \fr{2}{9} p^4    + 4 q^4   + \fr{2}{9} p^2 q^2 
                         + \biggl( \fr{2}{9} p^2   + \fr{4}{3} q^2  \biggr) k^2 \biggr] h(p) \Phi(q) \biggr\}                             
                 \label{4th nonlinear term in momentum space}        
\end{eqnarray}
and
\begin{eqnarray}
    && M_\P^2 e^{2\hphi} \fr{k^2}{(2\pi)^2} \int^\infty_\lambda 
         \fr{dp}{p^2} \fr{dq}{q^2} \theta(k-|p-q|)\theta(p+q-k) 
          \biggl\{   6 \Phi(p) \pd_\eta^2 \Phi(q) 
                 \nonumber \\
    && 
          + 6 \pd_\eta^2 \Phi(p) \Phi(q)
          + 6 \pd_\eta \Phi(p) \pd_\eta \Phi(q)
          + 12 \pd_\eta \hphi  \, \Phi(p) \pd_\eta \Phi(q) 
          + 12 \pd_\eta \hphi  \, \pd_\eta \Phi(p) \Phi(q) 
                 \nonumber \\
    && 
          + 12 \bigl( \pd_\eta^2 \hphi + \pd_\eta \hphi \pd_\eta \hphi \bigr) \Phi(p) \Phi(q)
          + 3 \bigl( p^2 +q^2 + k^2 \bigr) \Phi(p) \Phi(q) 
          - 8 h(p) \pd_\eta^2 \Phi(q)         
                \nonumber \\
    &&
          + 4 \pd_\eta h(p) \pd_\eta \Phi(q)
         - 16 \pd_\eta \hphi \, h(p) \pd_\eta \Phi(q)      
         + 8 \pd_\eta \hphi  \, \pd_\eta h(p) \Phi(q)
                 \nonumber \\  
    &&
         - 16 \bigl( \pd_\eta^2 \hphi + \pd_\eta \hphi \pd_\eta \hphi \bigr) h(p) \Phi(q)
         - \biggl( \fr{2}{3} p^2   + 14 q^2   +  2 k^2 \biggr)  h(p) \Phi(q)     \biggr\}   .              
                 \label{2nd nonlinear term in momentum space}        
\end{eqnarray}
Here, the momentum integral in (\ref{2nd nonlinear term in momentum space}) has no problem on upper bound due to the strength of its negative power. On the other hand, the absence of problems in (\ref{4th nonlinear term in momentum space}) derived from the fourth-derivative Riegert action will be guaranteed by the property that the traceless tensor field becomes smaller in the high-momentum region.\footnote{
When considering higher momentum regions, it will be necessary to introduce momentum dependence in the running coupling constant such as ${\bar t}^2(\tau,k) = \{ \b_0 \log[ \tau_\Lam^2 (1/\tau^2 + k^2/a(\tau)^2)] \}^{-1}$, even though the $k$-dependence disappears as soon as inflation begins.
} 

\section{Simplification of Nonlinear Terms and Numerical Evaluation}
\setcounter{equation}{0}

\noindent
In order to numerically solve the above system of nonlinear evolution equations, we have to describe the momentum integrals by sums and rewrite it to a multiple simultaneous differential equation in which many fields with different momentum are linked. However, it is not easy to actually solve such a multiple differential equation. Here we further simplify the equations and examine them to evaluate how the nonlinear term specifically affects patterns of the spectrum. At that time, we do not care too much about overall amplitude accuracy.

First of all, since we cannot solve the evolution equation if two time variables are mixed, we rewrite the conformal time $\eta$ to the proper time $\tau$. Using the scale factor $a$, the Hubble variable $H$, and its time derivatives $Y=\pd_\tau H$ and $Z=\pd_\tau^2 H$, the differential operators with respect to $\eta$ are rewritten as
\begin{eqnarray*}
  \pd_\eta &=& a \pd_\tau, 
            \nonumber \\
  \pd_\eta^2 &=& a^2 D_\tau^2,  \quad D_\tau^2 =  \pd_\tau^2 + H \pd_\tau, 
        \nonumber \\
  \pd_\eta^3 &=& a^3 D_\tau^3,  \quad D_\tau^3 = \pd_\tau^3 +3 H \pd_\tau^2 
                 + \bigl( Y  + 2 H^2 \bigr) \pd_\tau, 
        \nonumber \\
  \pd_\eta^4 &=& a^4 D_\tau^4 , \quad
             D_\tau^4  = \pd_\tau^4 +6 H \pd_\tau^3 + \bigl( 4 Y +11 H^2 \bigr) \pd_\tau^2 
                             +\bigl( Z + 7H Y + 6 H^3 \bigr) \pd_\tau .
\end{eqnarray*}
The inflationary variables are also rewritten as
\begin{eqnarray*}
  \pd_\eta \hphi &=& a H,   \quad
  \pd_\eta^2 \hphi = a^2 \bigl( Y + H^2 \bigr), \quad
  \pd_\eta^3 \hphi = a^3 \bigl( Z  +4 H Y + 2H^3 \bigr), 
        \nonumber \\
  \pd_\eta^4 \hphi &=& a^4 \bigl( \pd_\tau Z +7 H Z +4 Y^2 +18 H^2 Y +6 H^4 \bigr) ,
\end{eqnarray*}
where the last $\pd_\eta^4 \hphi$ was used when deriving (\ref{homogeneous equation in proper time}).

Decompose the scale factor as $a(t)=a(t_i) \times {\bar a}(t)$, where $t \, (=H_\D \tau)$ is the dimensionless time as introduced before and ${\bar a}(t_i) = 1$. Then, a dimensionless quantity obtained by dividing the physical momentum squared $k^2/a^2$ by $H_\D^2$ can be expressed using the comoving Planck mass $m \,$(\ref{comoving energy scales}) as
\begin{eqnarray}
      P_a (t,k) =\fr{k^2}{H_\D^2 a^2(\tau)} = \fr{k^2}{m^2 {\bar a}^2(t)}  .
             \label{physical momentum squared}
\end{eqnarray}
Similarly, we introduce a dimensionless differential operator defined by $D_t^n = D_\tau^n/H_\D^n = \pd_t^n + \cdots ~(n=2,3,4)$, in which the Hubble variables are replaced with normalized $\barH = H/H_\D$, $\barY = Y/H_\D^2$, and $\barZ = Z/H_\D^3$.

Next, we simplify the nonlinear terms as follows. First, applying an approximation that the field correlation vanishes instantaneously when distance exceeds the correlation length $\xi_\Lam$, let the field $f(k)$ be a variable with values such as a step function for $k \geq \lam$. We then focus on the following function that arises in the Fourier transforms of the field products:
\begin{eqnarray}
     \fr{1}{(2\pi)^2} \fr{k^2}{p^2 q^2} \theta(k-|p-q|)\theta(p+q-k) .
         \label{integrand in nonlinear term}
\end{eqnarray}
When displaying this function in the physical range $p, \, q \geq \lam$, there are two sharp peaks at $(p,q)=(\lam, k)$ and $(k,\lam)$. Therefore, we extract these neighborhoods and evaluate them. Supposing that the field variable $f(k)$ is the one averaged over a narrow range $\Delta$ around the momentum $k$, and considering only the contribution from the narrow region near the peak where $dpdq$ is $\Delta^2$, we simplify (\ref{fourier transform of product fg}) as follows:
\begin{eqnarray}
    (fg)(k) = \fr{1}{(2\pi)^2} \fr{\Delta^2}{\lam^2} \bigl[ f(\lam) g(k)  + f(k) g(\lam) \bigr] .
            \label{simplified form of nonlinear term}
\end{eqnarray}
We also simplify (\ref{fourier transform of product dfdg}) in the same way. The evolution equation derived in this way is written down in Appendix B, in which $TR_{[\lam]}(k)=0 \,$(\ref{fluctr equation}) is a simplified version of the trace equation (\ref{final trace equation}) while $SP(k)= 0 \,$(\ref{flucsp equation}) is the constraint equation (\ref{final constraint equation}).

Hence, we have to solve a system of nonlinear evolution equations consisting of the four equations
\begin{eqnarray}
   TR_{[\lam]}(k)=0, \quad  SP(k)=0,  \quad TR_{[\lam]}(\lam)=0,  \quad  SP(\lam)=0 ,
           \label{simplified coupled evolution equation}
\end{eqnarray}
simultaneously, together with the homogeneous equation (\ref{homogeneous equation in proper time}). 
Here note that the last two $TR_{[\lam]}(\lam)=0$ and $SP(\lam)=0$ form a system of equations representing the nonlinear evolution of the field with momentum $\lam$, and can be solved with only these two equations. In this way, although originally we have to solve a multi-line coupled equation system consisting of many fields with different momentums, it is now reduced to a two-line system consisting only of the fields with momenta $k$ and $\lam$, which can be solved relatively easily.

\begin{figure}[h]
\begin{center}
\includegraphics[scale=0.32]{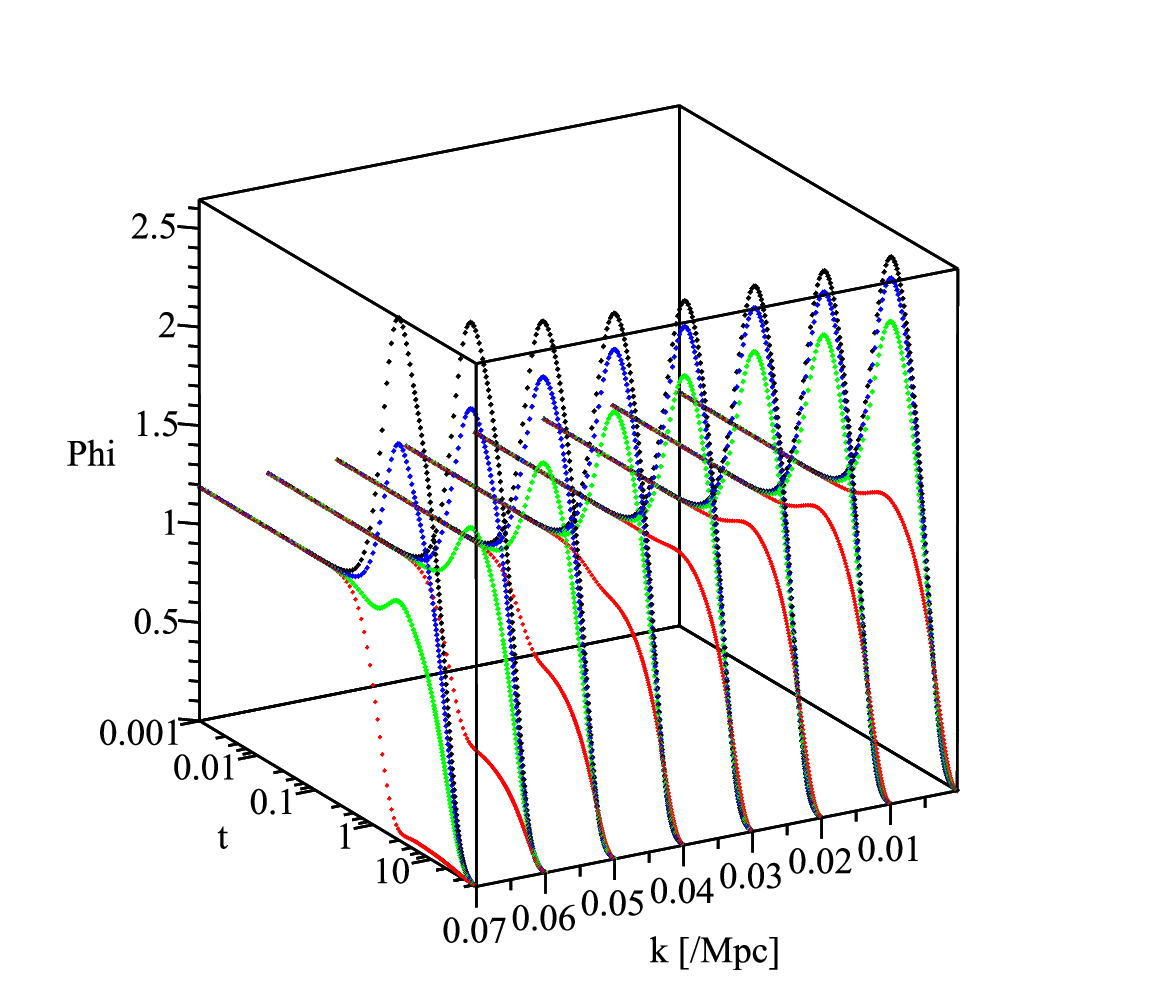}
\includegraphics[scale=0.32]{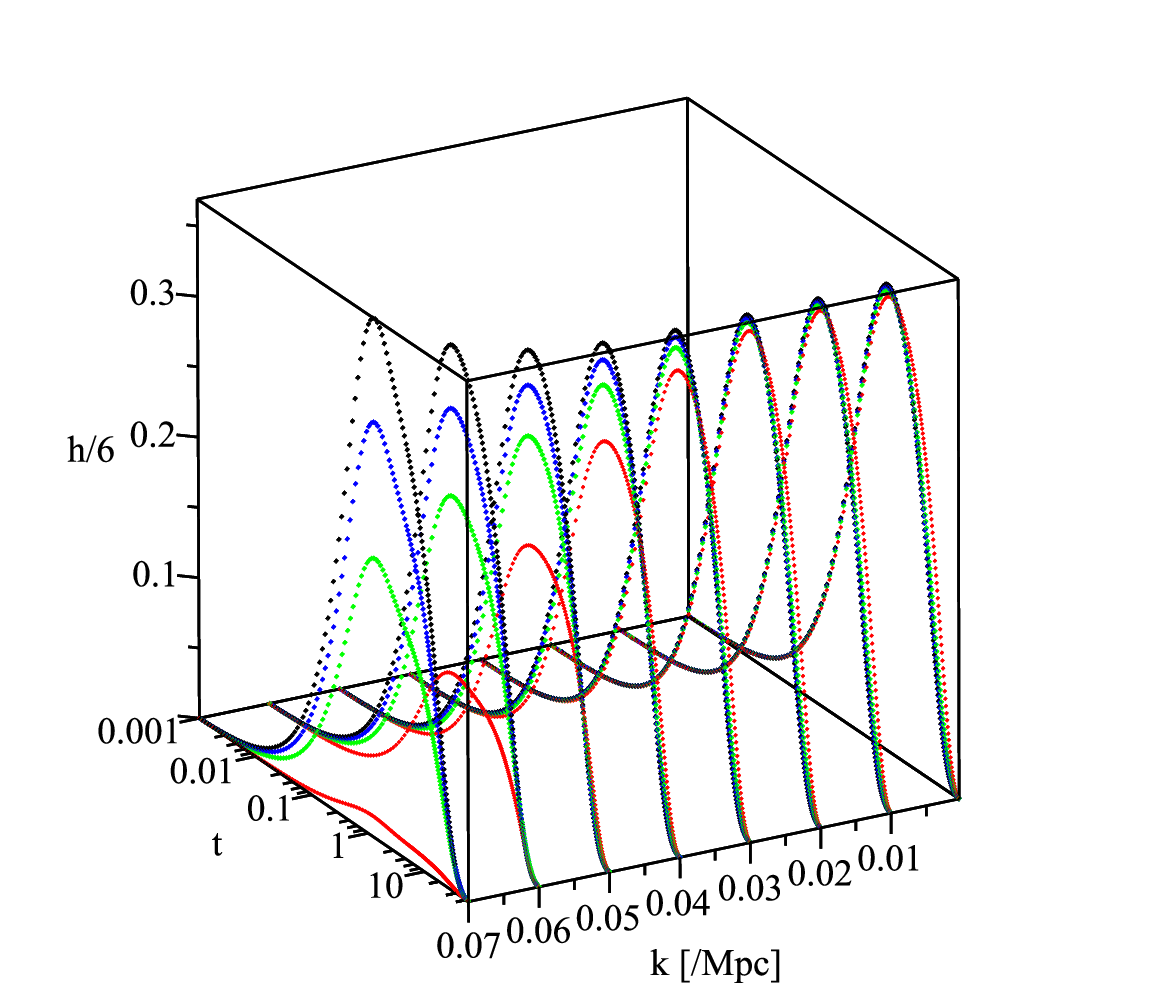}
\end{center}
\caption{\label{evolution of Phi and h}{\small 
Time evolution of the fluctuations including the nonlinear effects for $b_c=7$ and $m=0.02\, {\rm Mpc}^{-1}$. 
The end line denotes the phase transition time set at $t_\Lam =60$. Each shows $\Phi$ and the $h$-component in $\Phi$, respectively. Here, $\Delta/2\pi$ is $1.3\lam$ (black), $\lam$ (blue), and $0.7\lam$ (green) from the top, and the red is the result for the linear equation ($\Delta=0$), where the black is already shown in Figs.\ref{evolution of Phi in log and normal time} and \ref{last stage and primordial spectrum}. }}
\end{figure}

Let us solve this simplified evolution equation to see how the nonlinear terms work. Here $\Delta$ is introduced as a phenomenological parameter that effectively expresses the strength of the nonlinear term, and its value is assumed to be approximately $\Delta/2\pi \sim \lam$ because the coefficient of (\ref{simplified form of nonlinear term}) becomes about $1$.

In solving the nonlinear evolution equation (\ref{simplified coupled evolution equation}), we have to pay attention to the role of $SP=0$. As mentioned near the end of Section 3, this equation acts like a constraint condition connecting the initial stage where the running coupling constant $\bar{t}$ is small until the moment of the spacetime phase transition where it diverges. It indicates that the universe at $\bar{t}^2 \to 0$ before inflation is in a state where only the fluctuation of $\Phi = \vphi$ with $h=0$ exists. On the other hand, at the dynamical time $t_\Lam$ when $\bar{t}^2$ diverges, the fluctuation changes to that satisfying $\Phi =h/3$.

\begin{figure}[h]
\begin{center}
\includegraphics[scale=0.28]{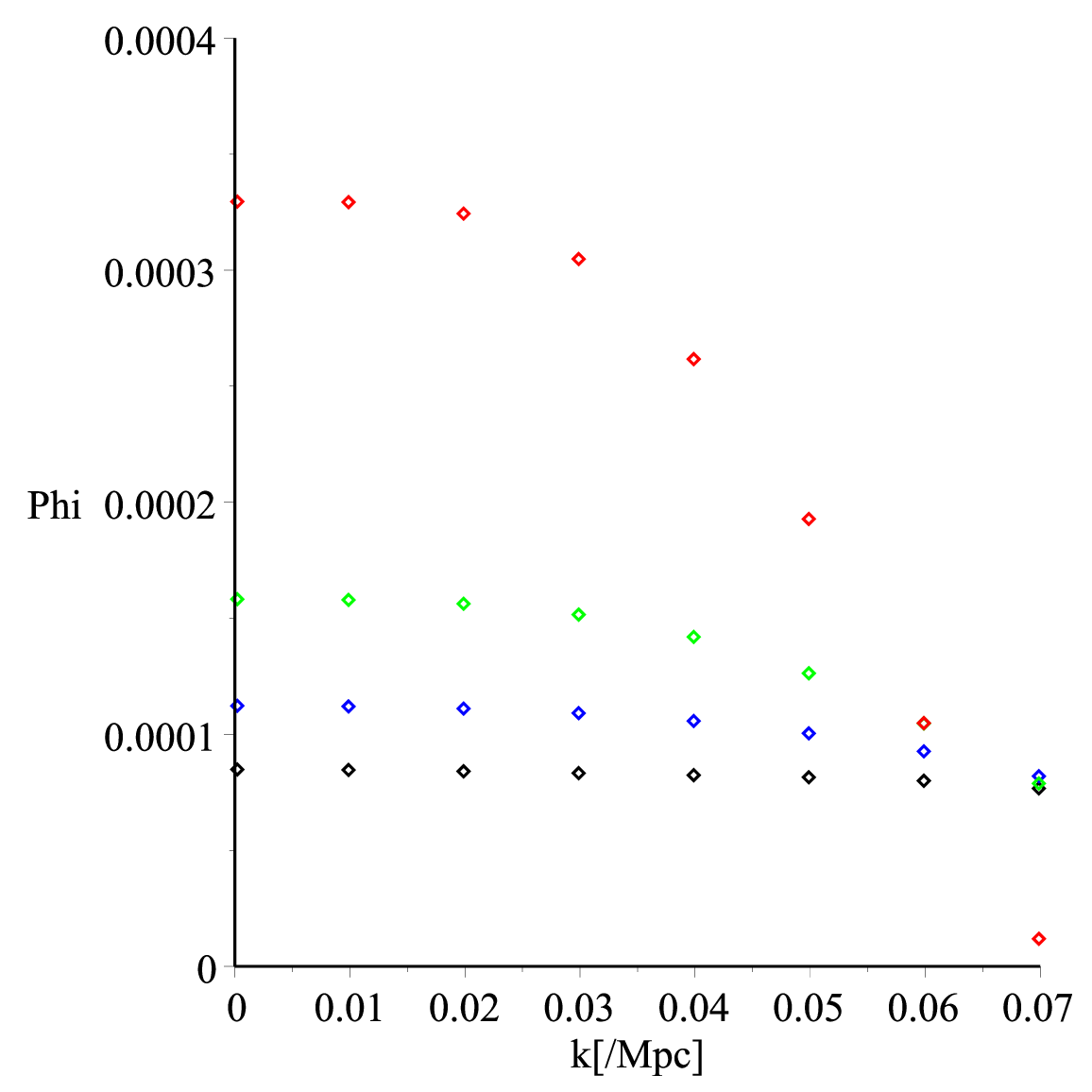}
\includegraphics[scale=0.28]{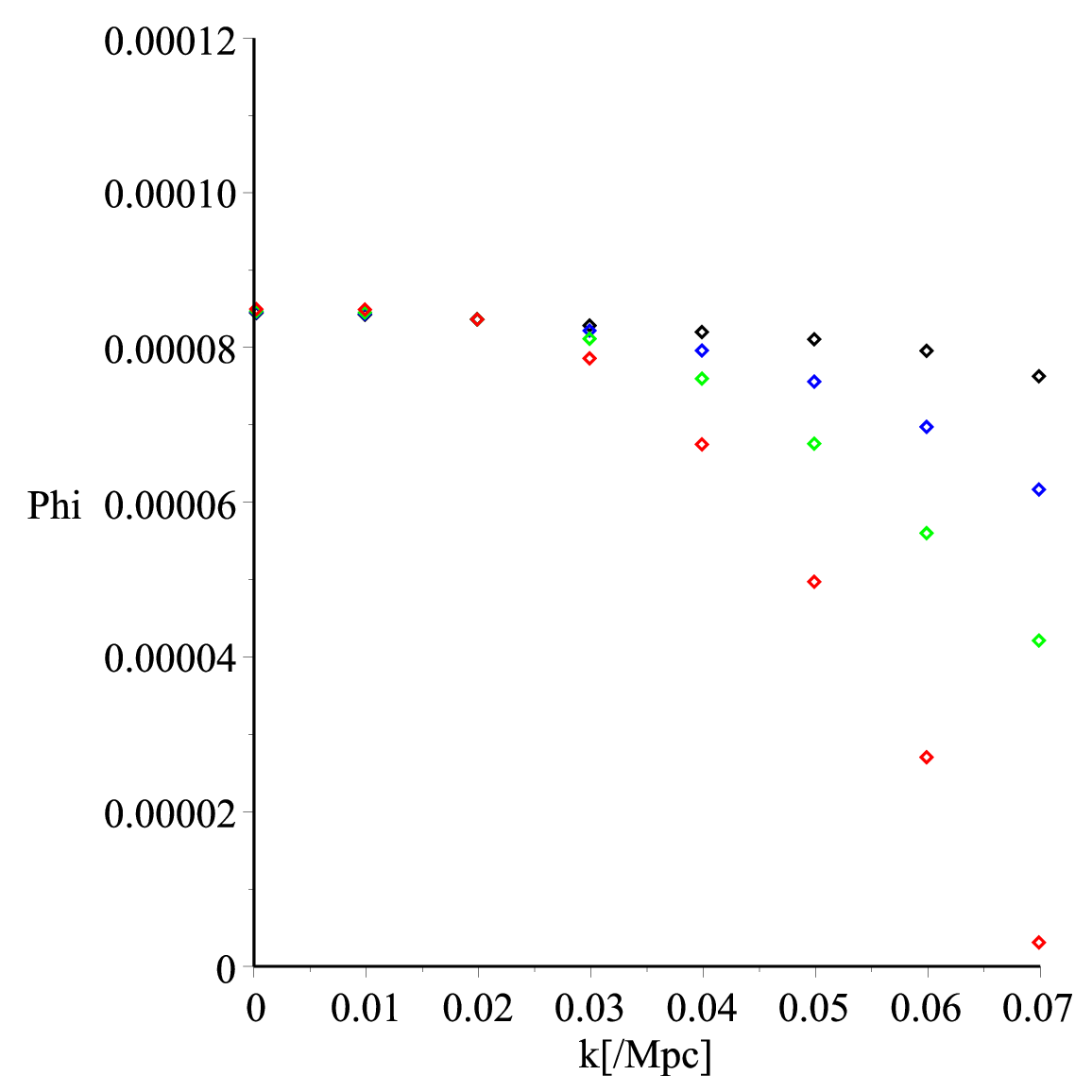}
\end{center}
\caption{\label{primordial spectra}{\small Plots of the value of $\Phi$ at the time of the spacetime phase transition. The right is the one normalized so that the others match the black at $k=m$. It can be seen that the nonlinear term serves to maintain the scale invariance of the spectrum beyond the comoving Planck scale $m$.}}
\end{figure}

To solve the evolution equation while preserving the constraint $SP=0$, we need to solve it as a boundary value problem. The initial time $t_i$ is set to $10^{-3}$, well before the Planck time, and the amplitude of $\Phi$ at that time is given by the square root of its scale-invariant power spectrum $\langle |\vphi(t_i, k)|^2 \rangle=\pi^2/b_c$ obtained by Fourier transform of the two-point correlation function (\ref{two point function}) \cite{hhy10}. Thus the initial value of $\Phi$ is set to
\begin{eqnarray}
      \Phi(t_i,k) = \fr{\pi}{\sq{b_c}}  
            \label{initial value of Phi}
\end{eqnarray}
and $\pd_t \Phi(t_i,k) = \pd_t^2 \Phi(t_i,k) = \pd_t^3 \Phi(t_i,k) =0$ for $k \geq \lam$. In order to set the boundary condition on the field $h$, we rewrite it as $h = 3 (\Phi - X)$ and impose the boundary condition on $X$. Since the initial value of $h$ is zero, $X(t_i,k)$ is equal to (\ref{initial value of Phi}), while $X$ should vanish at the dynamical time $t_\Lam (=60)$, thus we set
\begin{eqnarray}
       X(t_i,k) = \fr{\pi}{\sq{b_c}},  \qquad   X(t_\Lam, k) =  0 .
            \label{boundary condition}
\end{eqnarray}
To evaluate time evolution of the field with momentum $k$, we have to solve (\ref{simplified coupled evolution equation}) with the boundary conditions (\ref{initial value of Phi}) and (\ref{boundary condition}) at $k$, together with those where $k$ is replaced with $\lam$. Further comments on how to solve it are summarized in Appendix B.

The fluctuation which was scale-invariant before inflation deforms and eventually reduce with time. 
Fig.\ref{evolution of Phi and h} shows the results of time evolution of the fluctuations solved in this way for various $\Delta$, in which one of the results (black) is already displayed in Fig.\ref{evolution of Phi in log and normal time} and Fig.\ref{last stage and primordial spectrum} together with an enlarged view near the phase transition. The line of $k=\lam$ on the right end of each figure is the result of solving only the latter two equations in (\ref{simplified coupled evolution equation}), while each $k$ line is obtained by solving the four equations simultaneously.

Fig.\ref{primordial spectra} plots the value of $\Phi$ at the time of the phase transition, and the square of it  will give the primordial spectrum that is an initial condition of the subsequent time evolution. The right normalized panel shows that the nonlinear term has effects of flattening the spectrum even in the region $k > m$. In the previous linear equation results \cite{hhy06}, the spectrum was greatly deformed at $k > m$, but it can be seen that the nonlinear effect suppresses the deformation up to the high momentum region, exceeding 3 times $m$.

Here, we examined the evolution equation simplified by introducing the parameter $\Delta$, but in fact it will be necessary to solve a system of evolution equations consisting of a very large number of the fields obtained by describing the momentum integrals of (\ref{4th nonlinear term in momentum space}) and (\ref{2nd nonlinear term in momentum space}) by sums. In addition, although only nonlinear terms up to second order in the fields are considered here, it is believed that higher-order nonlinear terms contribute in higher momentum regions. In fact, here the linear approximation exhibits large undulations in the spectrum and some strange behavior, which suggests that the nonlinear effects given by the exponential function become essential in that region.

Why, then, do the nonlinear terms with the exponential factor play an important role in maintaining the scale invariance despite the presence of mass scale? The Einstein-Hilbert action having such a factor is one of conformal fields in terms of conformal field theory. It is generally believed that  correlation functions in the theory with it as a potential term exhibit  a power-law behavior. It may be considered that this behavior is manifested.

Thus, we can expect that the scale invariance before inflation is maintained up to the relatively-high momentum region due to the nonlinear effects, and is conveyed to the present CMB. However, to what extent will such an effect be maintained? Could unexpected structures emerge in regions of even higher momentum \cite{hss}? The results here are the first step toward clarifying these matters.

\begin{center}
{\bf Acknowledgements}
\end{center}

I wish to thank Shinichi Horata for his great contributions to the numerical calculations in the early stages of this research.

\appendix

\section{Energy-Momentum Tensor for Each Sector}
\setcounter{equation}{0}

The energy-momentum tensor for each sector defined by (\ref{definition of energy-momentum tensor}) is given here \cite{hhy06}.  In the following, $\Box = \pd^\lam \pd_\lam = - \pd_\eta^2 + \lap3$ is d'Alembertian in the flat background and $\chi_\mu = \pd_\lam h^\lam_{~\mu}$. The symmetric product is $a_{(\mu} b_{\nu)} = (a_\mu b_\nu + a_\nu b_\mu )/2$.

\paragraph{Riegert sector} 
\begin{eqnarray}
 \bT^{\rm R}_{\mu\nu}
  &=& \fr{b_c}{8\pi^2} B \biggl\{ 
  4\Box \phi \pd_\mu\pd_\nu \phi 
  -4 \pd_{(\mu}\Box\phi \pd_{\nu)}\phi 
  -\fr{8}{3}\pd_\mu\pd_\lam \phi \pd_\nu\pd^\lam \phi 
  +\fr{4}{3}\pd_\mu \pd_\nu \pd_\lam \phi \pd^\lam \phi
 \nonumber \\ && 
  +\fr{2}{3}\pd_\mu \pd_\nu \Box \phi 
  +\eta_{\mu\nu} \left[ -\Box\phi \Box\phi 
  +\fr{2}{3} \pd_\lam \Box \phi \pd^\lam \phi 
  +\fr{2}{3} \pd_\lam \pd_\s \phi \pd^\lam \pd^\s \phi 
  -\fr{2}{3} \Box^2 \phi \right] 
 \nonumber \\ && 
  -4 \pd_{(\mu}h^\lam_{\nu)} \Box\phi \pd_\lam \phi 
  +2 \pd^\lam h_{\mu\nu} \Box\phi \pd_\lam \phi 
  -4 h^{\lam\s} \pd_\lam \pd_\s \phi \pd_\mu\pd_\nu \phi 
  -4 \chi^\lam \pd_\mu \pd_\nu \phi \pd_\lam \phi
 \nonumber \\ && 
  +4 h^{\lam\s} \pd_\lam \pd_\s \pd_{(\mu}\phi \pd_{\nu)}\phi 
  +4 \pd_{(\mu}h^{\lam\s} \pd_\lam \pd_\s \phi \pd_{\nu)} \phi 
  +4\chi^\lam \pd_\lam \pd_{(\mu}\phi \pd_{\nu)}\phi 
 \nonumber \\ && 
  +\fr{8}{3} h^{\lam\s} \pd_\mu \pd_\lam \phi \pd_\nu \pd_\s \phi  
  +\fr{4}{3} \pd_{(\mu} h^{\lam\s} \pd_{\nu)} \pd_\lam \phi \pd_\s \phi 
  +4 \pd^\lam h^\s_{(\mu} \pd_{\nu)}\pd_\lam \phi \pd_\s \phi
 \nonumber \\ && 
  -4 \pd^\lam h^\s_{(\mu} \pd_{\nu)} \pd_\s \phi \pd_\lam \phi 
  -\fr{4}{3} h^{\lam\s} \pd_\mu \pd_\nu \pd_\lam \phi \pd_\s \phi 
  -\fr{4}{3} \pd_{(\mu}h^\lam_{\nu)} \pd_\lam \pd_\s \phi \pd^\s \phi
 \nonumber \\ &&
  +\fr{2}{3} \pd^\lam h_{\mu\nu} \pd_\lam \pd_\s \phi \pd^\s \phi 
  +\fr{4}{3} \pd_\mu \pd_\nu h^{\lam\s} \pd_\lam \phi \pd_\s \phi 
  -4 \pd^\lam \pd_{(\mu} h^\s_{\nu)} \pd_\lam \phi \pd_\s \phi
 \nonumber \\ &&
  +2 \pd^\lam \pd^\s h_{\mu\nu} \pd_\lam \phi \pd_\s \phi
  -4 \pd^\lam \chi_{(\mu} \pd_{\nu)} \phi \pd_\lam \phi
  +4 \Box h^\lam_{(\mu} \pd_{\nu)} \phi \pd_\lam \phi
 \nonumber \\ &&
  +\fr{4}{3} \pd_{(\mu} \chi_{\nu)} \pd_\lam \phi \pd^\lam \phi
  -\fr{2}{3} \Box h_{\mu\nu} \pd_\lam \phi \pd^\lam \phi 
  +\fr{4}{3} \pd_\lam \chi^\lam \pd_\mu \phi \pd_\nu \phi
 \nonumber \\ &&
  -4 h^\lam_{(\mu} \pd_{\nu)} \pd_\lam \phi \Box \phi
  +2 h^\lam_{(\mu} \pd_{\nu)} \Box \phi \pd_\lam \phi
  +2 h_{\lam (\mu} \pd^\lam \Box \phi \pd_{\nu)} \phi
 \nonumber \\ &&
  +\fr{8}{3} h^\lam_{(\mu} \pd_{\nu)} \pd^\s \phi \pd_\lam \pd_\s \phi  
  -\fr{4}{3} h^\lam_{(\mu} \pd_{\nu)} \pd_\lam \pd_\s \phi \pd^\s \phi
 \nonumber \\ &&
  + \eta_{\mu\nu} \biggl[ 
     2h^{\lam\s} \pd_\lam \pd_\s \phi \Box \phi 
     +2\chi^\lam \Box \phi \pd_\lam \phi 
     -\fr{2}{3}h^{\lam\s} \pd_\lam \pd_\s \pd_\rho \phi \pd^\rho \phi 
 \nonumber \\ && \quad
     -\fr{2}{3}\chi^\lam \pd_\lam \pd_\s \phi \pd^\s \phi 
     + 2 \pd^\lam \chi^\s \pd_\lam \phi \pd_\s \phi 
     -\fr{2}{3} h^{\lam\s} \pd_\lam \Box \phi \pd_\s \phi
     -\fr{4}{3} h^{\lam\s} \pd_\lam \pd_\rho \phi \pd_\s \pd^\rho \phi
 \nonumber \\ && \qquad\quad
     -\fr{4}{3} \pd^\rho h^{\lam\s} \pd_\rho \pd_\lam \phi \pd_\s \phi 
     -\fr{4}{3} \Box h^{\lam\s} \pd_\lam \phi \pd_\s \phi 
     -\fr{2}{3} \pd_\lam \chi^\lam \pd_\s \phi \pd^\s \phi  
     \biggr]
 \nonumber \\ &&
  -\fr{4}{3} \pd_{(\mu} h^{\lam\s} \pd_{\nu)} \pd_\lam \pd_\s \phi 
  -\fr{2}{3} h^{\lam\s} \pd_\lam \pd_\s \pd_\mu \pd_\nu \phi
  -\fr{8}{3} \pd_\mu \pd_\nu h^{\lam\s} \pd_\lam \pd_\s \phi
  -\fr{2}{3} \chi^\lam \pd_\lam \pd_\mu \pd_\nu \phi
 \nonumber \\ &&
  +\fr{8}{3} \pd_{(\mu} \chi^\lam \pd_{\nu)} \pd_\lam \phi 
  -\fr{2}{3} \pd_\mu \pd_\nu \chi^\lam \pd_\lam \phi
  -\fr{2}{3} \pd_{(\mu} h^\lam_{\nu)} \pd_\lam \Box \phi 
  +\fr{1}{3} \pd^\lam h_{\mu\nu} \pd_\lam \Box \phi
 \nonumber \\ &&
  +4 \pd^\lam \pd_{(\mu} h^\s_{\nu)} \pd_\lam \pd_\s \phi
  -2 \pd^\lam \pd^\s h_{\mu\nu} \pd_\lam \pd_\s \phi
  -\fr{14}{3} \pd_{(\mu} \chi_{\nu)} \Box \phi
  +\fr{7}{3} \Box h_{\mu\nu} \Box \phi
 \nonumber \\ &&
  -2 \pd_\lam \chi^\lam \pd_\mu \pd_\nu \phi
  +4 \pd_\lam \chi_{(\mu} \pd_{\nu)} \pd^\lam \phi
  -4 \Box h^\lam_{(\mu} \pd_{\nu)} \pd_\lam \phi
  +\fr{2}{3} \pd_\lam \pd_{(\mu} \chi^\lam \pd_{\nu)} \phi
 \nonumber \\ &&
  -\fr{2}{3} h^\lam_{(\mu} \pd_{\nu)} \pd_\lam \Box \phi
 \nonumber \\ &&
  + \eta_{\mu\nu} \biggl[ 
       \fr{4}{3} h^{\lam\s} \pd_\lam \pd_\s \Box \phi 
       +\fr{4}{3} \pd^\rho h^{\lam\s} \pd_\rho \pd_\lam \pd_\s \phi
       +\fr{4}{3} \chi^\lam \pd_\lam \Box \phi
       +\fr{8}{3} \Box h^{\lam\s} \pd_\lam \pd_\s \phi
 \nonumber \\ && \qquad\quad
      -\fr{8}{3} \pd^\lam \chi^\s \pd_\lam \pd_\s \phi 
      +\fr{2}{3} \Box \chi^\lam \pd_\lam \phi
      +2 \pd_\lam \chi^\lam \Box \phi
      -\fr{1}{3} \pd^\lam \pd_\s \chi^\s \pd_\lam \phi
       \biggr] 
 \nonumber \\ &&
  -\fr{1}{9} \pd_\mu \pd_\nu \pd_\lam \chi^\lam
  +\fr{1}{9} \eta_{\mu\nu} \Box \pd_\lam \chi^\lam   \biggr\} 
           \label{Riegert energy-momentum tensor}
\end{eqnarray}
and the trace is 
\begin{eqnarray}
    \bT^{{\rm R}\mu}_{~~~\mu} 
      &=& \fr{b_c}{8\pi^2} B \biggl\{
                 -2 \Box^2 \phi   + 4 h^{\mu\nu} \Box \pd_\mu \pd_\nu \phi
                 + 4 \pd^\lam h^{\mu\nu} \pd_\lam \pd_\mu \pd_\nu \phi
                 + 4 \chi^\lam \pd_\lam \Box \phi
                      \nonumber \\
       &&
                + 4 \Box h^{\mu\nu} \pd_\mu \pd_\nu \phi      + 2 \Box \chi^\lam \pd_\lam \phi
                + \fr{4}{3} \pd_\lam \chi^\lam \Box \phi         - \fr{2}{3} \pd^\lam \pd_\s \chi^\s \pd_\lam \phi
                + \fr{1}{3} \Box \pd_\lam \chi^\lam  \biggr\}  .   
                      \nonumber \\
      &&           \label{trace of Riegert sector}         
\end{eqnarray}

\paragraph{Weyl sector}
\begin{eqnarray}
      \bT^{\rm W}_{\mu\nu} 
        =   - \fr{2}{t^2} \biggl\{ \Box^2 h_{\mu\nu}     -  2 \Box \pd_{(\mu} \chi_{\nu)}  
               + \fr{2}{3} \pd_\mu \pd_\nu  \pd_\lam \chi^\lam    
               + \fr{1}{3} \eta_{\mu\nu} \Box \pd_\lam \chi^\lam     \biggr\}  
                \label{Weyl energy-momentum tensor}
\end{eqnarray}
and the trace vanishes.

\paragraph{Einstein-Hilbert sector}
\begin{eqnarray}
      \bT^{\rm EH}_{\mu\nu} &=& M_{\rm P}^2 \e^{2\phi} \biggl\{ 
        2 \pd_\mu \pd_\nu \phi                    -2 \pd_\mu \phi \pd_\nu \phi
        +\eta_{\mu\nu} \bigl( -2 \Box \phi -\pd^\lam \phi \pd_\lam \phi \bigr)
        -\pd_{(\mu} \chi_{\nu)}                             
                    \nonumber \\ 
      &&
         +\half \Box h_{\mu\nu}        -2 h^\lam_{(\mu} \pd_{\nu)} \pd_\lam \phi   
         +2 h^\lam_{(\mu} \pd_{\nu)} \phi \pd_\lam \phi
         -2 \pd_{(\mu} h^\lam_{\nu)} \pd_\lam \phi   +\pd^\lam h_{\mu\nu} \pd_\lam \phi
                     \nonumber \\
      &&
         +\eta_{\mu\nu} \biggl( \half \pd_\lam \chi^\lam       +2 h^{\lam\s} \pd_\lam \pd_\s \phi
                        +h^{\lam\s} \pd_\lam \phi \pd_\s \phi   +2 \chi^\lam \pd_\lam \phi    \biggr)
         \biggr\} 
              \label{Einstein-Hilbert energy-momentum tensor}
\end{eqnarray}
and the trace is 
\begin{eqnarray}
    \bT^{{\rm EH}\mu}_{~~~~\mu} 
      &=&  M_\P^2  \, e^{2\phi} \Bigl\{  
                 - 6 \Box \phi         - 6 \pd^\lam \phi \pd_\lam \phi 
                 + \pd_\lam \chi^\lam         + 6 h^{\lam\s}  \pd_\lam \pd_\s \phi 
                 + 6 \chi^\lam \pd_\lam \phi
                         \nonumber \\
      && 
                  + 6 h^{\lam\s} \pd_\lam \phi \pd_\s \phi  \Bigr\} .
               \label{trace of Einstein-Hilbert sector}
\end{eqnarray}

\section{Summary of Simplified Nonlinear Evolution Equation and How To Solve It}
\setcounter{equation}{0}

\noindent
Time evolution of the fluctuations is obtained by simultaneously solving the evolution equation of the fluctuations and the homogeneous equation that determines the inflationary background.

The homogeneous equation can be expressed as
\begin{eqnarray}
    \pd_t \bar{Z} +7 \bar{H} \bar{Z} +4 \bar{Y}^2 +18 \bar{H}^2 \bar{Y} +6 \bar{H}^4 
    - 3 \biggl[ 1 + \fr{\gm_1}{\kappa} \bar{\a}_G(t)  \biggr]^\kappa \bigl( \bar{Y} + 2 \bar{H}^2 \bigr) = 0 ,
          \label{normalized homogeneous equation}
\end{eqnarray}
using the dimensionless time $t=H_\D \tau$ and the normalized Hubble variables, $\bar{H}=\pd_t \bar{a}/\bar{a }$, $\bar{Y}=\pd_t \bar{H}$, and $\bar{Z}=\pd_t \bar{Y}$, defined in Section 5. Here, the time-dependent running coupling constant is denoted as $\bar{\a}_G(t)=\bar{t}^2(t)/4\pi$. The initial time $t_i = H_\D/E_i$ is set to $10^{-3}$ so that $E_i$ sufficiently larger than the Planck energy. Numerical calculations are stopped at $t_\Lam-\eps$ just before the running coupling constant diverges, and $\eps$ is reduced until the result does not change.
Fig.\ref{scale factor and H} is the result calculated in this way.

In the following, the evolution equation of fluctuations that we actually dealt with and the method for solving it are described in detail. We use the trace equation (\ref{final trace equation}) with multiplied by $(8\pi^2/b_c)\times (1/\bar{B}) \times (1/H_\D^4 a^4 )$ and simplified the nonlinear terms as in Sections 4 and 5. Denoting it as $TR_{[\lam]}(k)=0$ yields:
\begin{eqnarray}
   && TR_{[\lam]}(k) =
         -2 D_t^4 \Phi(k)           - 4P_a(k) D_t^2 \Phi(k)   
         - 2 P_a^2(k) \Phi(k)      - \fr{4}{3} \barH D_t^3 h(k)   
                        \nonumber \\
   &&\quad
         -\fr{16}{3} \bigl( \barY + \barH^2  \bigr) D_t^2 h(k)       + \fr{4}{9} P_a(k) D_t^2 h(k)
         - 8 \bigl( \barZ + 4 \barH \barY + 2 \barH^3  \bigr) \pd_t h(k)
                        \nonumber \\
   &&\quad
         - \fr{20}{9} \barH P_a(k)  \pd_t h(k)      - \fr{32}{9} \bigl( \barY + \barH^2  \bigr) P_a(k) h(k)     
         + \fr{4}{9} P_a^2(k)  h(k)    
                         \nonumber \\ 
   &&\quad
         + \fr{1}{(2\pi)^2} \fr{\Delta^2}{\lam^2} \biggl\{
                - 8 \pd_t h(\lam) D_t^3 \Phi(k)    - 8 \pd_t h(k) D_t^3 \Phi(\lam)
                - \fr{16}{3} D_t^2 h(\lam) D_t^2 \Phi(k)       
                         \nonumber \\
   &&\quad
                - \fr{16}{3} D_t^2 h(k) D_t^2 \Phi(\lam)   
                - \fr{4}{3} D_t^3 h(\lam) \pd_t \Phi(k)         - \fr{4}{3}  D_t^3 h(k) \pd_t \Phi(\lam)
                         \nonumber \\
   &&\quad
                + \biggl[  - \fr{20}{9} P_a(\lam)    + \fr{16}{3} P_a(k)  \biggr] h(\lam) D_t^2 \Phi(k)                
                + \biggl[  - \fr{32}{9} P_a(k)    + \fr{20}{3} P_a(\lam)   \biggr] h(k) D_t^2 \Phi(\lam)
                          \nonumber \\
   &&\quad
                + \biggl[ - \fr{20}{9} P_a(\lam)   - \fr{8}{3} P_a(k)  \biggr] \pd_t h(\lam) \pd_t \Phi(k)                       
                + \biggl[ - \fr{20}{9} P_a(k)   - \fr{8}{3} P_a(\lam)  \biggr] \pd_t h(k) \pd_t \Phi(\lam)                       
                          \nonumber \\
   &&\quad
               - \fr{2}{3} P_a(\lam) D_t^2 h(\lam) \Phi(k)     - \fr{2}{3} P_a(\lam) D_t^2 h(k) \Phi(\lam)
                          \nonumber \\
   &&\quad
              + \biggl[ - \fr{2}{9} P_a^2(\lam)     + \fr{16}{3} P_a^2(k)   
                                             + \fr{4}{9} P_a(\lam) P_a(k) \biggr] h(\lam) \Phi(k)                                               
                          \nonumber \\
   &&\quad
              + \biggl[  4 P_a^2(\lam)    + \fr{14}{9} P_a(\lam) P_a(k)  \biggr] h(k) \Phi(\lam)    \biggr\}                                           
                          \nonumber \\
   &&\quad
         + \biggl[ 1 + \fr{\gm_1}{\kappa}\bar{\a}_G(t) \biggr]^\kappa \biggl\{  
            6 D_t^2 \Phi(k)      + 12 \barH \pd_t \Phi(k) 
            +  12 \bigl(  \barY  + 2 \barH^2 \bigr) \Phi(k)     
            + 6 P_a(k) \Phi(k)
                          \nonumber \\
   &&\quad
               + 4 \barH  \pd_t h(k)         - 8 \bigl( \barY + 2 \barH^2 \bigr) h(k)     
               - \fr{4}{3} P_a(k) h(k)  \biggr\}
                         \nonumber \\
   &&\quad
       +  \fr{1}{(2\pi)^2} \fr{\Delta^2}{\lam^2} \biggl[ 1 + \fr{\gm_1}{\kappa}\bar{\a}_G(t) \biggr]^\kappa  
          \biggl\{
           12 \Phi(\lam) D_t^2 \Phi(k)      +  12 \Phi(k) D_t^2 \Phi(\lam)    
           +  12 \pd_t \Phi(\lam) \pd_t \Phi(k)      
                         \nonumber \\
   &&\quad
           + 24 \barH \bigl[ \Phi(\lam) \pd_t \Phi(k)  +  \Phi(k) \pd_t \Phi(\lam)  \bigr]                          
           + 24 \bigl( \barY + 2 \barH^2 \bigr) \Phi(\lam) \Phi(k)
                         \nonumber \\
   &&\quad
           + \bigl[ 12 P_a(k)    + 6 P_a(\lam) \bigr] \Phi(\lam) \Phi(k)
           - 8 h(\lam) D_t^2 \Phi(k)      - 8 h(k) D_t^2 \Phi(\lam)
                         \nonumber \\
   &&\quad
           + 4 \pd_t h(\lam) \pd_t \Phi(k)            + 4 \pd_t h(k) \pd_t \Phi(\lam)  
           - 16 \barH \Bigl[ h(\lam) \pd_t \Phi(k)  + h(k) \pd_t \Phi(\lam)  \Bigr]      
                         \nonumber \\
   &&\quad                        
           + 8 \barH \bigl[ \pd_t h(\lam) \Phi(k)   +  \pd_t h(k) \Phi(\lam)   \bigr]   
           - 16 \bigl( \barY  + 2 \barH^2 \bigr) \bigl[  h(\lam) \Phi(k)   +  h(k) \Phi(\lam) \bigr]                  
                       \nonumber \\
   && \quad
           - \biggl[ \fr{2}{3} P_a(\lam)    + 16 P_a(k)  \biggr] h(\lam) \Phi(k)
           - \biggl[ \fr{8}{3} P_a(k)    + 14 P_a(\lam) \biggr] h(k)  \Phi(\lam)    \biggr\}  .
                \label{fluctr equation}                              
\end{eqnarray}
If writing the constraint equation (\ref{final constraint equation}) multiplied by $(3/16) \times \bar{t}^2 \times (1/H_\D^2 a^2)$ as $SP(k)=0$, then we get
\begin{eqnarray}
   && SP(k) =
         D_t^2 h(k)   + \fr{1}{3} P_a(k) h(k)    
          +  \fr{b_c}{8\pi} \bar{\a}_G(t) \bar{B}(t) \biggl\{
                D_t^2 \Phi(k)        + 2 \barH \pd_t \Phi(k)  
                     \nonumber \\
   &&  
                + 6 \bigl( \barY +  \barH^2 \bigr) \Phi(k)   +  P_a(k) \Phi(k) 
                + \fr{2}{3} \barH \pd_t h(k)
                + \fr{2}{3} \bigl( \barY  - \barH^2 \bigr) h(k)  
                       \nonumber \\ 
   &&                    
               - \fr{2}{9} P_a(k) h(k)   \biggr\}
               +  \fr{b_c}{8\pi} \bar{\a}_G(t) \Bigl\{
               - 3 \Phi(k)  +  h(k)    \Bigr\} .
                 \label{flucsp equation}
\end{eqnarray}
Moreover, using (\ref{third derivative of h}), the terms with $D_t^3 h \, (= \pd_\eta^3 h/a^3 H_\D^3)$ in (\ref{fluctr equation}) should be rewritten to terms up to the third time-derivative of $\Phi$ and up to the second time-derivative of $h$ as
\begin{eqnarray*}
     D_t^3 h (k)  &=& 
         - \fr{1}{3} P_a (t) \pd_t h(k)  
          + \fr{\pd_t {\bar \a}_G(t)}{{\bar \a}_G(t)}  \biggl( D_t^2 h + \fr{1}{3} P_a(k) h \biggl)
                         \nonumber \\
         && - \fr{b_c}{8\pi} {\bar \a}_G(t) \biggl\{  
                  \bar{B}(t)  \biggl[  D_t^3 \Phi (k) + 2 \barH D_t^2 \Phi (k)    
                 + 8 \bigl( \barY + \barH^2 \bigr) \pd_t \Phi (k)
                         \nonumber \\
          &&\qquad
                 + 6 \bigl(\barZ + 4 \barH \barY + 2 \barH^3 \bigr) \Phi(k)     +  P_a(k) \pd_t \Phi(k) 
                 + \fr{2}{3} \barH D_t^2 h(k)       
                         \nonumber \\
          &&\qquad
                 +  \fr{4}{3} \barY \pd_t h(k)       + \fr{2}{3} \bigl(  \barZ   - 2 \barH^3  \bigr) h(k)     
                 - \fr{2}{9} P_a(k) \pd_t h(k)   \biggr]    
                        \nonumber \\
          &&\qquad  
                + \pd_t \bar{B} (t) \bigr)    \biggl[  D_t^2 \Phi(k)  + 2 \barH \pd_t \Phi (k)   
                + 6 \bigl( \barY + \barH^2 \bigr)  \Phi (k)  
                        \nonumber \\
          &&\qquad
                +  P_a (k) \Phi (k)   + \fr{2}{3} \barH \pd_t h(k)  
                + \fr{2}{3} \bigl( \barY  - \barH^2 \bigr)  h (k)    - \fr{2}{9} P_a (k) h (k)  \biggr]     \biggr\}
                        \nonumber \\
          &&  + \fr{b_c}{8\pi} \bar{\a}_G (t)  \Bigl\{   3 \pd_t \Phi (k)   +  6 \barH \Phi (k)    
                        - \pd_t h (k)   - 2 \barH h (k)  \Bigr\} .
\end{eqnarray*}

The time evolution of the field with momentum $k$ including the nonlinear effects is obtained by simultaneously solving the four equations $TR_{[\lam]}(k)=0$, $SP(k)=0$, $TR_{[\lam]}(\lam )=0$, and $SP(\lam)=0$. Here, note that the last two form a closed system of equations for the field with momentum $\lam$ that can be solved by them alone.

When solving (\ref{fluctr equation}) and (\ref{flucsp equation}), we have to solve the homogeneous equation simultaneously to determine the inflationary background, but the scale factor ${\bar a}$ appears only in the form $P_a(k) \,$(\ref{physical momentum squared}) in these equations. Since the solution of ${\bar a}$ in the early stage is given by $\bar{a}(t)=e^{\barH_i (t-t_i)} $ with $\barH_i = 1/\sq{\bar{B}(t_i)}$ and also $P_a(k)$ vanishes rapidly with time, using this function to reduce computational load does not affect the result.

Here, in order to further reduce computational load, we fix the inflationary background as $\barH =\barH_i$, $\barY=0$, and $\barZ=0$. If contributions from the nonlinear terms are small, the evolution equation can be solved without such fixing. However, this manipulation only raises or lowers overall amplitude of the spectrum and does not affect the spectral pattern, which is almost determined in the early stage. Therefore, the background is fixed here so that the calculations can be performed with more aggressive parameter values.

Using the ratio of the two physical scales (\ref{ratio of two physical scales}), the running coupling constant is written as $\bar{\a}_G(t )=[4\pi \b_0 \log (N^2/t^2)]^{-1}$. When performing the numerical calculation actually, we first convert this to $\bar{\a}_G^c(t)=[4\pi \b_0 \log ((N+c)^2/t^2)]^{-1 }$ and find a solution of $c=1$, then use it as an approximate solution to find the solution of $c=0$. However, since the introduction of $c$ also only raises and lowers the amplitude slightly as a whole and does not affect the spectral pattern, we here present only the results for c=1 to reduce the load, as mentioned above. In this case, there is no need to introduce $\eps$.

Special programs are needed to handle boundary value problems for differential equations where some variables that are not observable diverge at boundaries \cite{bvpsolver}. Moreover, we need to solve simultaneous equations with multiple variables. The commercially available Maple software has a built-in program for such boundary value problems, and we use it here. Specifically, we use the ``dsolve'' command in Maple15, set it to ``method=bvp[midrich], abserr=1e-2, initmesh=1000, maxmesh=8192 (maximum value)'', and solve it numerically. In addition, as $\Delta$ gets larger, it becomes more difficult to converge initial Newton iterations. In that case, we use the ``continuation'' parameter to start from a smaller $\Delta$ where the initial iteration converges, leading to the desired result.

Figs.\ref{evolution of Phi in log and normal time}, \ref{last stage and primordial spectrum}, \ref{evolution of Phi and h}, and \ref{primordial spectra} are the results of solving the nonlinear evolution equations with the boundary conditions (\ref{initial value of Phi}) and (\ref{boundary condition}), by simplifying some functions and calculation methods, as described above. In order to perform large-scale computations involving many fields with different momentums with higher precision, a special program such as the Fortran software, BVP\_SOLVER \cite{bvpsolver}, and a device that can run it will be necessary.


\end{document}